\documentclass[11pt,a4paper]{article}

\usepackage{graphicx}
\usepackage{eurosym}
\usepackage{amssymb}
\usepackage{amsmath}

\begin{document}

\newcommand{\nsms}{$n$\,m$^{-2}$\,s$^{-1}$}
\newcommand{\nps}{$n$\,s$^{-1}$}
\newcommand{\musvh}{$\mu$Sv\,h$^{-1}$}
\newcommand{\msvh}{mSv\,h$^{-1}$}

\title{Simulated Fast Neutron Instrument Backgrounds at a MW Spallation Source}

\author{P. M. Bentley}
\date{}
\maketitle

\begin{abstract} 
Like every megaproject, the European Spallation Source (ESS) currently
under construction in Lund, Sweden, has had to make a number of
inevitable compromises to remain within budget and on schedule.  Some of these
compromises carry technical consequences with a potential impact on the operational characteristics
of the facility and the neutron scattering instruments in particular.  In order to prepare for operation of the facility, it is timely to consider again the subjects of instrument backgrounds and shielding from the project design phase.  The expected ranges of instrument backgrounds, and the relative strengths of each contributor, are estimated using multiple physics simulations.  These are compared with other leading spallation sources.  Finally, several suggestions are given to reduce the background contributions in each case, should they prove to be dominant in practice when the commissioning phase of the facility begins.
\end{abstract}

\section{Introduction}
The European Spallation Source (ESS) is intended to be eventually the
most powerful source of neutrons available for scientific research.  If operational upgrades are completed, it will approximately match the world-leading Institut Laue-Langevin
(Grenoble, France) for time-averaged neutron flux, and exceed
the instantaneous neutron brightness of the pulsed Japan Proton
Accelerator Research Complex (J-PARC).

During the conceptual design phase of the facility, an extensive public document summarised the concepts upon which the facility was founded \cite{ESS-TDR}, drawing upon the wealth of experience from existing spallation and reactor sources.  This was in order to establish broad technical standards, lay the groundwork for defining best current practice, and identify major technical challenges.  Subsequent work defined in more detail some appropriate standards and methods for the optical and shielding components in the user-facing instrumentation \cite{NOSG-HANDBOOK}.   Half of the cost of spallation source instruments fall into the bracket of neutron optics and shielding, and these two technical areas are strongly coupled.  Important common factors are performance, particularly signal-to-noise ratios; cost and correct strategic planning \cite{BENTLEY-COST-INTERNAL-REPORT,BENTLEY-COST-OPTIMISATION}; the required simulation expertise; and the activation of components, particularly how it affects the maintainability of instruments, licensing, and nuclear waste.

Some of these efforts to standardise on best practice were
successful, but it is inevitable that non-technical factors drive other definitions of optimality and success
and influence the direction of the project.  
Many of these deviations from technical best practice will have no
bearing upon the operation of the facility, but others could affect it
in a significant way.  In order to prepare for this work, it is
therefore timely to provide a technical perspective on these challenges, so that
the future work is anticipated and planned to the degree that such an
ambitious facility deserves.

This article will begin with a summary of the knowledge established early in the project, and identify key risks along with the origins of many challenges.  Simulations and measurements will then be presented in support of the technical standards, and the identification of mitigation steps that can be taken if these risk scenarios arise.

\section{Technical Overview}

\begin{figure}
  \includegraphics[width=0.9\textwidth]{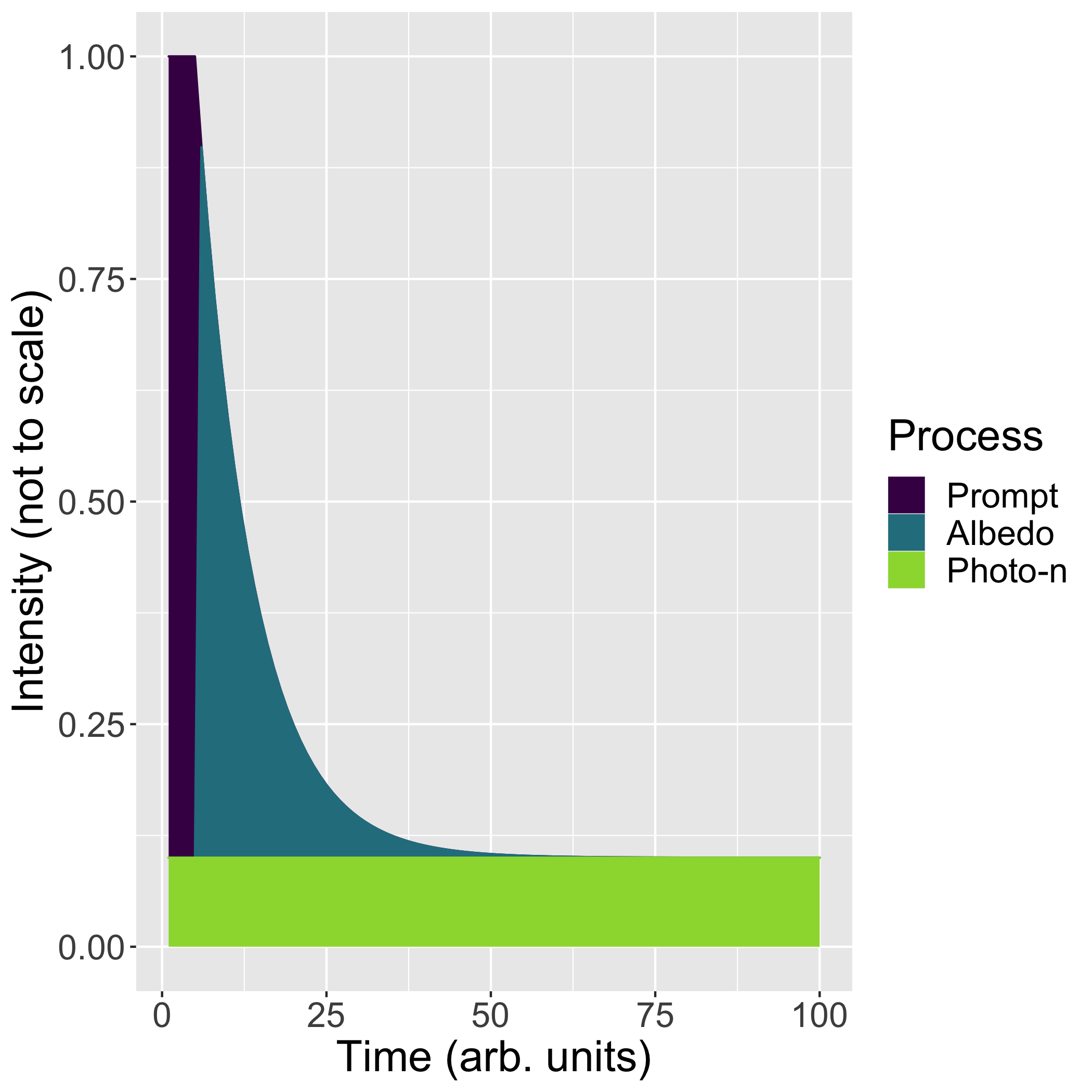}
  \caption{Illustration of fast neutron background components at a pulsed spallation source.  Only a single measurement frame is shown, the next pulse occurs at $time=100$.}
  \label{fig:BGCartoon}
\end{figure}

An illustration, not to scale, of a fast neutron instrument background at a pulsed spallation source instrument is shown in figure \ref{fig:BGCartoon}.
There we see three effects that are summarised.  The first, labelled ``prompt'', represents the highest energy neutrons, from MeV all the way up to the incident proton beam energy, whilst the proton beam illuminates the target.  The long tail behaviour, labelled ``albedo'' is caused by fast neutrons experiencing multiple scattering events on equipment, the atmosphere, and the ground, which in turn generate more secondary particles of lower energy.  Some of these are neutrons that thermalise near the detector and are counted, and others may create high energy effects around the detector that trigger as a false positive event, i.e. resembling thermal neutron capture.  Finally, photonuclear processes generate an almost flat, continuous source of MeV neutrons.

There are of course photoneutrons generated from $\gamma$-illumination as part of the spallation process, but their contribution to the prompt and albedo parts of the curve is almost negligible.  The flat contribution discussed here is mainly from radioactive decay, illuminating materials in which neutrons are loosely bound.  Photonuclear reactions are an efficient source of neutrons.  Indeed, there are several projects to build facilities based on this physics instead of spallation sources (e.g. \cite{LUE-200}) --- this should give pause for thought as to how intense photonuclear background sources can be.

There are other processes that contribute to a flat, time-independent background.  Examples include cosmic rays, electronic signal noise, impurities in the detector equipment, etc.

With assistance from collaborators, fast neutron backgrounds at existing facilities were surveyed over a number of years, and a comparison between each beamline was begun.  These mainly included the instruments HYSPEC and CNCS \cite{SNS-SPECTROMETERS,CNCS} at the SNS, Oak Ridge, Tennessee, USA; The SINQ facility at PSI in Villigen, Switzerland; and LET \cite{BEWLEY_LET} at RAL, Oxfordshire, UK, but additionally OFFSPEC --- also at RAL.  To a lesser degree, we also looked at AMATERAS at JPARC in Japan.  These will be discussed in more detail shortly.

During the fact-finding period around 2012, our visits to some of these existing spallation source facilities provided direct contact with their technical experts to identify the main design risks for a new facility of this kind.  Moreover, the ESS is a long-pulse source, which is a unique concept \cite{MEZEI-LONG-PULSE}, and this has some additional importance.  Our hosts reported four key technical risks, based on hands-on experience, for the ESS project.  These are:
\begin{description}
  \item[Backgrounds] --- in other words instrument noise --- due to both the long pulse length and unprecedented
    source power;
    \item[Optical alignment] --- whilst $>$100\,m long guides exist at
      reactors, the long average guide length in combination with the
      large amounts of shielding needed by spallation sources
      increases the risk of misalignment, which would reduce instrument performance below designed parameters;
    \item[Activation] due to the large time-averaged flux of the long
      pulse in combination with high energy physics.  Some parts of ESS will greatly exceed familiar levels of
      activation as experienced at reactors and short pulse sources.  This impacts on maintenance, safety, and waste.  The continuous, MW-power SINQ facility is perhaps the closest existing precedent in this area.
    \item[Material longevity] again due to the source power and
      particle energies involved.
\end{description}

The risks are listed in order of relative importance as defined by the
consulted experts.  This article will concentrate on the shielding aspects, the optics may be considered later in a subsequent report.

Some of the greatest shielding challenges at ESS have already been described in an earlier article \cite{COPPER-GUIDES} and will be only briefly summarised here.  The first challenge is that the ESS has the thinnest bulk shielding design around the target when compared to other leading spallation sources around the world, whilst at the same time it is also intended to be the world's most powerful source of neutrons.  Given that the legal limits for radiation safety are roughly similar between these facilities, there is an obvious inconsistency.

The second challenge was a number of technical aspects related to instrument backgrounds in detail.
\begin{itemize}
\item Steel shielding is transparent at high energy, and contrary to some beliefs the addition of extra carbon to create ``high carbon'' steels does not block the resonant windows sufficiently.  Stainless steels do a much better job, and were employed in the ESS target for their corrosion resistance properties, but elsewhere the cost and activation problems of stainless steel are prohibitive.  This drives us to consider alternatives --- see \cite{COPPER-GUIDES} and references therein.

\item Fast neutron backgrounds cannot be eliminated by accumulating over longer measurement times, and arguments comparing pulsed fast neutrons with those from natural background phenomena are erroneous --- these both confuse systematic error with statistical error.  The time-dependent prompt part, and its long tail behaviour, is a systematic error.  On the other hand, the time-independent photo-nuclear component, and natural backgrounds, are often statistical errors when integrated over typical measurement times.

\item Only partial success is achieved by subtracting fast neutron backgrounds from the data.  This method relies on constancy of the background level between measurements.  The reflectometry scientists at SNS verbally reported positive outcomes.  On the other hand, for spectroscopy there are documented observations of the relative background level varying as a function of time and causing significant issues in scientific productivity \cite{SNS-BACKGROUND-SURVEY}.

\item Finally, it cannot be assumed that any region in the data that is affected by fast neutron backgrounds can be ``vetoed'' or removed --- this assumes that the long tail behaviour of the background does not encroach significantly on the measurement region of interest for an experiment.  This assumption is almost certainly invalid, especially for a long pulse source.  Even at short pulse sources, many of our collaborators reported significant issues on this point.
\end{itemize}


A third challenge is that both background and activation effects are caused by physical
processes that do not occur at significant rates at nuclear reactor
sources.  It is almost instinctive for
neutron scattering scientists to cover equipment with an absorber,
such as boron, to reduce neutron activation, since this is the optimal solution for thermal or cold neutron beams.  It has almost no effect for steel and heavy
concrete shielding in direct view of the spallation source.  The vast
majority ($>$90\%) of the worker dose rate for spallation-exposed iron
comes from $^{54}$Mn, which is created via the reaction $^{54}$Fe($n,p$)$^{54}$Mn.  The cross section for this reaction is significant, $\sim$100 mBarn, across incident energies of 30-100 MeV \cite{MICHEL-ACTIVATION}.  The absorption cross section of boron at MeV neutron energies and above is
negligible \cite{IAEA-XSECT-ATLAS}.  It only takes a few minutes of extra work to see this, by adding an activation tally to the shielding simulation (e.g. DCHAIN in PHITS \cite{PHITS}).

However, boron liners on shielding could be needed to
reduce argon activation in the air, which is from thermal neutron capture --- fast neutrons enter the shielding, thermalise, and then radiate back into the open spaces --- but that is a separate problem for establishing ventilation requirements and ensuring safety.  Those thermalising neutrons can also be a strong contribution to backgrounds if the shielding is part of the instrument.  

To reduce the activation of the bulk shielding itself, which affects maintenance, one must use carefully designed laminates that pay attention to these high energy physics processes.  One example is using a layer of several cm of lead on the front face of the steel shielding to attenuate the higher energy neutrons as they enter (it also reduces gamma shine to the workers from the activated shielding), and mix boron into the deeper material layers to capture the thermalising neutrons.  This laminate concept for high energy shielding represents unfamiliar territory to many of those with working experience at reactor-based neutron sources.  There remain a few examples of laminates that were retained in ESS components, but the bulk of shielding projects did not favour laminate options.

On background, the key lessons learned were reported in 2012
\cite{SNS-LESSONS-LEARNED} at the Spallation Neutron Source (SNS, at
Oak Ridge, Tennessee, USA), and these are quoted here:
\begin{enumerate}
\item Address background needs early
\item Instrument shielding structures a.k.a. ``caves'', create a
  ``sea'' of neutrons and the structures themselves act as moderators
\item All components along the length of the beamline have a potential
  influence on the background
\item Components on the cave walls will need to be shielded
\item Instrument design review must be held with a specific charge to
  ask if sufficient background shielding has been made part of the
  design
\end{enumerate}

Modelling the background to take such effects into account is often quite
different from similar work employed in radiological safety.  In
safety work, no credit can be taken for equipment in the beam unless
that equipment is under some form of safety control, for example attaching physical keys and locks into critical safety items that are tied into the beam control systems.  This is rarely done at scale due to the expenses involved, and usually applies only to doors, beam shutters, and a small number of critical shielding blocks.  Since the two scenarios of safety and background are different, then two variance reduction
approaches need to be used.  In safety one typically wishes to drive the simulation through bulk shielding and quantify the low fluxes on the far side of a shield wall.  In background work, one might be interested mainly in the neutrons that are back-scattered from the wall, down long channels, or via specific physics channels that are such low probability to be irrelevant for safety but highly significant to background effects.

\begin{figure}
  \includegraphics[width=0.9\textwidth]{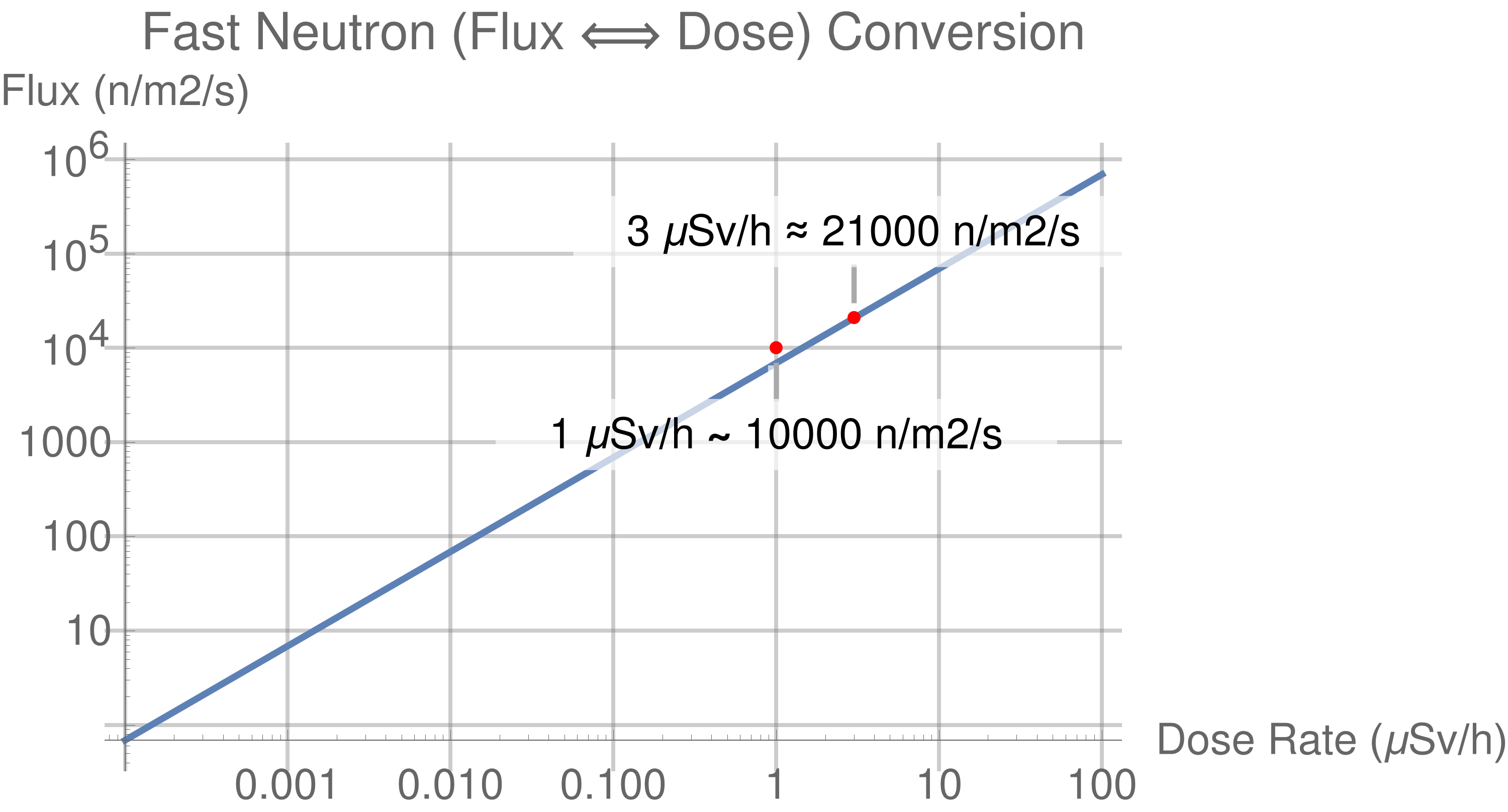}
  \caption{Dose rate to flux conversion for fast neutrons ($\sim
    1$~MeV) according to Sullivan \cite{SULLIVAN}.  1~$\mu$Sv/h
    is easily remembered as being roughly 10$^4$~\nsms, but the accurate
    number is around 7000~\nsms.}
  \label{fig:Dose2Flux}
\end{figure}

One can start to visualise the importance of the background problem by considering the fast neutron flux ($\gtrsim$1\,MeV)
for a given dose rate, assuming that all equipment is designed to meet
safety requirements.  The majority of neutrons emerging from thick shielding walls are at such energies (see figure \ref{fig:whyNotHeavyConcrete} for example).   Sullivan \cite{SULLIVAN} relates fast neutron fluence to dose rate at roughly $40\times10^{-15}$ Sv per
neutron.m$^{-2}$, from which it is fairly straight forward to show
that a contact dose rate of 3~$\mu$Sv/h corresponds roughly to 21,000
\nsms as shown in figure \ref{fig:Dose2Flux}.  This is the legal limit for some classes of workers in
European countries.  Many people remember 1~$\mu$Sv/h $\sim$ 10$^4$
n/m$^2$/s, though this is slightly less accurate it still provides an order of magnitude estimate.
Thus we can see that a moderately-sized instrument shielding wall, for example 3$\times$3\,m$^2$, may be radiating 10$^5$\,n/s if it is illuminated by fast neutrons and it is designed on the limit of safety.  Compared to the desired background count rates on some instruments, $\ll$1\,n/s \cite{NOSG-HANDBOOK}, these fast neutron leakage figures are so large they provoke immediate disbelief.  Thus we reach the ``golden rule'' of fast neutron backgrounds: \textbf{designing shielding purely for safety will result in sources of background that could be orders of magnitude higher than the scientific requirements of a functioning instrument}.  At PSI, for example, the contact dose rate generally around the facility is below 0.3\,$\mu$Sv/h, because radiation levels higher than this create a situation where instruments are unusable \cite{UWE-PRIVATE-COMM}.  This background level is an order of magnitude more stringent than the safety requirements, and this is a continuous source where backgrounds are essentially random errors that are more easily subtracted.  At a pulsed source, where backgrounds are partially systematic errors, the requirements exceed those at PSI.  The precise fast neutron flux that can be tolerated by an instrument must be determined by simulation.

This golden rule applies to two areas in particular: short instruments that are located nearby
the spallation target ($\lesssim$30\,m), and any wall and roof areas of the accelerator, target, and short instrument shielding that are a potential source of skyshine to more distant instruments ($>$100~m).  Instruments may have to contend with both effects as will be shown shortly.

During operations, the SNS produced a study summarising their efforts
to trace the source of fast neutron backgrounds on the HYSPEC
instrument \cite{SNS-BACKGROUND-SURVEY}, and after a decade of
work this represents the most complete published understanding
of these phenomena.

The count rates experienced on the SNS spectrometers HYSPEC and
CNCS associated with fast neutron backgrounds are around 1-4
neutrons per minute per metre of detector tube, and each tube measures
around 2-3\,cm in diameter.  This corresponds to roughly 1~\nsms.
One's first impression is that this is a low background rate, but the documented scientific requirements for several instrument classes are a $10^6$ signal-to-noise ratio \cite{NOSG-HANDBOOK}, and 1~\nsms exceeds this level by roughly two orders of magnitude on these spectrometers.

\begin{figure}
  \includegraphics[width=0.9\textwidth]{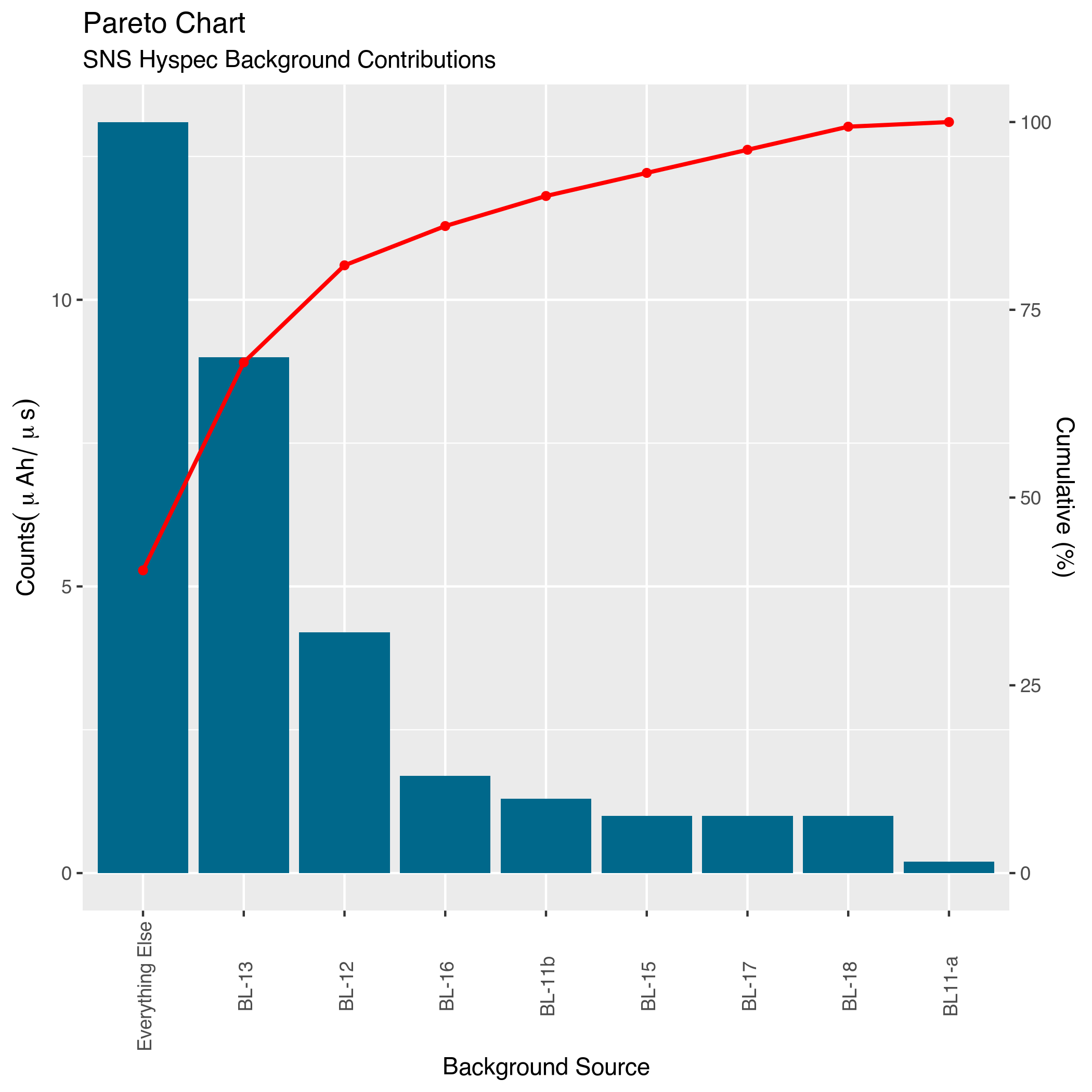}
  \caption{Pareto chart of fast neutron background sources at the
    SNS.}
  \label{fig:SNS-PARETO}
\end{figure}

The potential sources that were identified in the SNS study are
summarised in the Pareto chart shown in figure \ref{fig:SNS-PARETO}.
This data analysis ultimately quantifies all identifiable
neighbouring beamlines and the neutrons coming down the instrument's
own guide.  However, it shows that the largest source of background
remains unknown, and this unknown contribution represents more than
1/3 of the total background level.  Even if all the identifiable
sources of background could be addressed, at great expense, the
background would be reduced by around 60\% --- not even one order of
magnitude.

In our own earlier studies of the SNS \cite{SNS-BACKGROUND-SURVEY} and
PSI \cite{PSI-TARGET-NEUTRONICS}, it was found that, with the
exception of a small number of notable hotspots, most shielding
achieved safety requirements.

How these effects together could all be contributing to the background
levels for future ESS instruments will be explored in the next section.  Beginning in
section \ref{sec:caveResponses}, the response of the instrument cave is established, for a number of simple, fast-neutron illumination
scenarios.  

The overall balance of these quantified sources will be
estimated for the ESS for long and short instruments, based on the
best possible knowledge available during the facility design phase.  It will begin with a treatment of what the background may look like assuming most of the background comes down the instrument's own beam optics, including both random and systematic effects --- equivalent to the second largest bar in figure \ref{fig:SNS-PARETO}.  In later sections, additional external contributions to systematic errors will be explored, equivalent to the largest bar in figure \ref{fig:SNS-PARETO}, and an attempt to break that bar down further for ESS will be made.  Note that the relative intensity of these two bars is impossible to estimate reliably for the ESS in a generic way, since it depends so much on the design details of the instruments.

\section{Simulations and Results}

\subsection{Simulation Details}
There are many simulations that are presented in the following sections.  Since the reader may be interested in creating similar results for their own projects, and knowing that simulations can be rather time consuming, some technical details and working time will be reported.  The simulations were mostly performed on a Ryzen 7 2700 8-core workstation with 16\,GB RAM, running Ubuntu Linux v.18.04.5\,{LTS}.  The simulations originally circulated in draft form at ESS meetings around 2014-2015 were performed with GEANT4 \cite{GEANT4} on a Lenovo Thinkpad T440s running Ubuntu v.12.04 LTS or v.14.04 LTS, and such GEANT4 results are labelled so in this article.  Much of that earlier work, however, was re-generated here with better statistics for publication using PHITS v.3.17 \cite{PHITS} linked to OpenMPI v.2.1.1.  Unless specifically indicated otherwise, the reader should assume PHITS was used as described above.


\subsection{Cave Response\label{sec:caveResponses}}

Before diving into the individual fast neutron mechanisms contributing to backgrounds, one should consider the response of the instrument cave shielding (or the building surrounding the instrument if there is no cave) in order to model the effects that the structures and detector have for a given incident neutron energy.  As an example, this article explores a simplified 3$\times$3$\times$3\,m$^3$ cave made from concrete and 60\,cm thick concrete walls and roof.  A 3 cm thick $^3$He volume at 10 atmospheres was added across the rear wall.  A T-YIELD tally was used to count the number of triton particles generated within the detector volume, with the entire set of simulations normalised ``per source'' particle.  Alpha yield from $^{10}$B detectors could equally be modelled if necessary, but for simplification and fair comparison the method of modelling triton yield in helium is probably sufficient for the present purposes.

There are three response functions to quantify.  The first is the response of the cave to a fast neutron that is transported down the guide channel by multiple albedo events into the cave and strikes the back wall (``albedo'').  The second is the response to a neutron striking the external front or side wall surfaces (``direct illumination'') and the third is a neutron striking the cave roof from above (``skyshine'').  Though this is called skyshine here, any tall structures and external building rooves that are illuminated will scatter neutrons back down towards the instrument in this way.

\begin{figure}
  \includegraphics[width=0.9\textwidth]{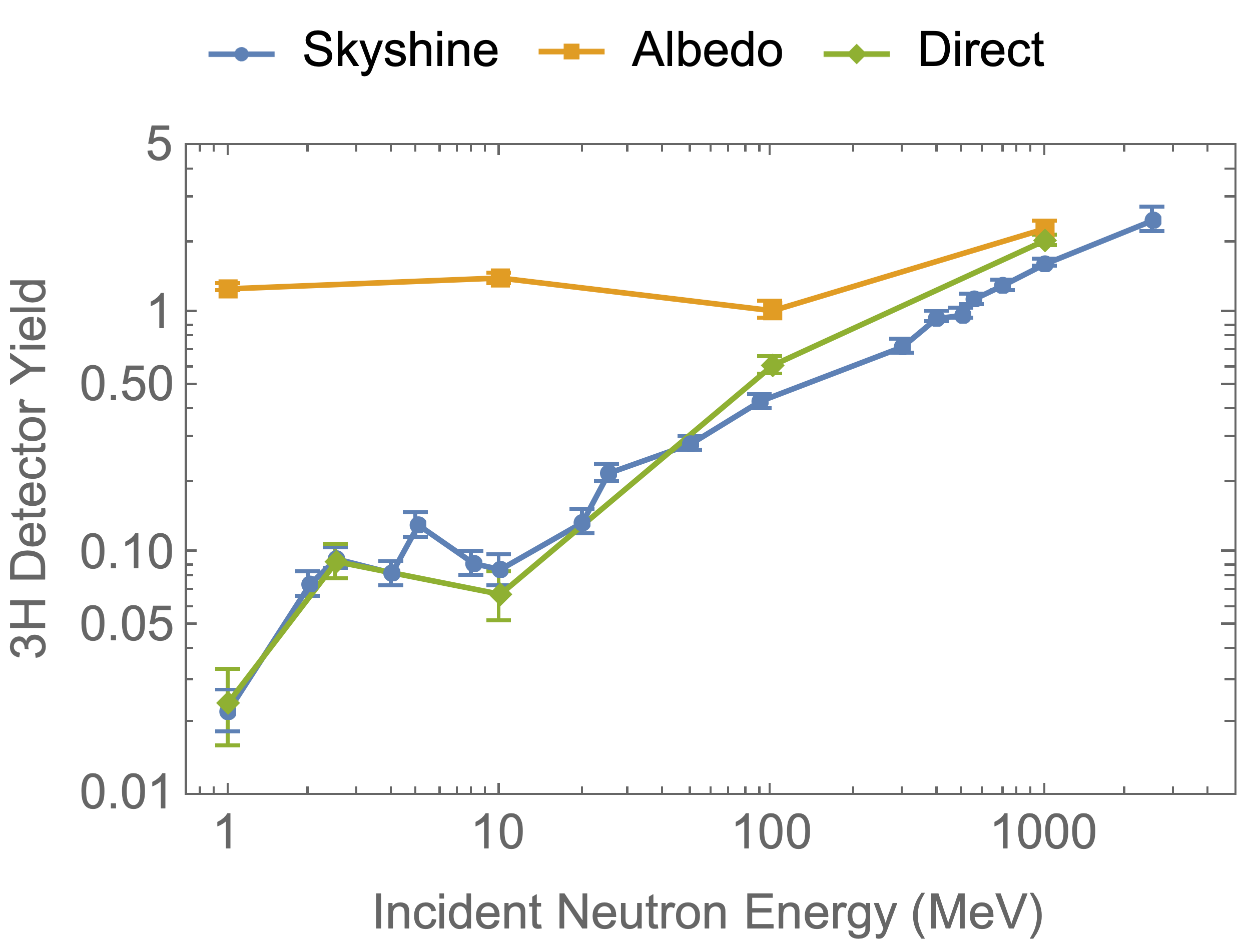}
  \caption{Energy dependent response of the cave and detector system to fast neutron illumination, with a simplified geometry as described in the text.}
  \label{fig:responseCurves}
\end{figure}

The results of these simulations are shown in figure \ref{fig:responseCurves}.
There one can see that neutrons emitted into the cave by albedo transport down the guide are essentially counted very efficiently across the full energy range, and one can ignore the response function.  For skyshine neutrons and direct illumination of the front of the cave, the response is almost a straight line on a log-log plot.  There are some visible resonance effects at lower energies, as expected.  These graphs will be referred to in subsequent sections when estimating the fast neutron background count rates from external cave sources.  Both skyshine and direct illumination attenuate at best down to a 2\% level but only at the lowest energies.  10\% is probably representative considering the spectral weight, for our order-of-magnitude estimates.  Many readers will have noticed that some of these data points are above 1, meaning that for each neutron entering the system more than one neutron is counted.  The reason this happens is the generation of secondary particles, which is an important part of spallation neutron target physics.

Each of the data points in figure \ref{fig:responseCurves} is an individual simulation, which took only a few minutes to compute.  Including building the geometry and organising the output data, the whole work is a few days and certainly less than two weeks.

\subsection{Albedo Transport}

Fast neutron transport \emph{via} albedo refers to the reflection of
neutrons at high energy, albeit in a stochastic manner.  Neutron supermirror guides, vacuum housings
and shielding act as conduits for fast neutrons that propagate quite
effectively if not carefully managed, and these are transported into the cave along with the desired cold and thermal neutrons.  Moreover, the timing of the arrival of these neutrons can encroach significantly into the measurement window for thermal/cold neutrons.

Measurements reported by Selph \cite{SELPH-FAST-NEUTRON-ALBEDO} reveal
a broad angular dependence of the fast neutron albedo on concrete,
which spans 30-60\%, with the most efficient scattering occurring at smaller
grazing angles.  A similar result happens at very small grazing angles to allow the transport of thermal neutrons by
supermirrors (for recent representative reflectivity curves, see for example \cite{SUPERMIRROR-REFLECTIVITIES}).  The main difference is that
supermirrors reflect only at the smallest grazing angles of a few degrees, but
fast neutron albedo reflectivity occurs at the perpendicular
90$^\circ$.  As a result, fast neutrons are quite easily transported
over large distances by neutron guides and surrounding infrastructure
if designed na\"{i}vely.

In addition to transport down the guide and shielding channels, neutron albedo
is expected at any sufficiently thick and dense surface.
Shielding walls themselves can deflect fast neutrons through large
angles, around corners, and over buildings.  A general
``rule-of-thumb'' that can be used with confidence is that half of the
fast neutrons shining on a surface will be scattered back.

The primary source of albedo transported background neutrons is the
spallation target itself, which may be modified in character by any shared shielding spaces between beamlines.

To quantify this albedo component for a long spectrometer at ESS, simulations were performed in a several stages.  The prompt, albedo transport of fast neutrons was done using a published source term \cite{ESS-SOURCE-TERM} and using the duct source variance reduction \cite{JPARC-CURVED-GUIDES} in PHITS for a 160\,m long channel.  The channel contains a 5$\times$5\,cm$^2$ glass guide channel with 1\,cm thick substrates.  The vacuum housings and shimming are modelled as a simple 1 cm thick rectangular cross section tube made from mild steel with 9$\times$9\,cm$^2$ internal dimensions.  The concrete shielding channel is 20$\times$20\,cm$^2$ surrounded by walls that are 50\,cm thick.  This is thinner than the expected shielding in reality, but the objective is to capture a representative albedo return and not to model the shielding attenuation.  

For simplicity, figure \ref{fig:responseCurves} allows us to assume a unity response to neutrons entering the instrument from the guide exit, and furthermore it is assumed that the temporal broadening effects within a $3\times 3\times 3$\,m$^3$ cave enclosure are negligible compared to those through the 160\,m long guide channel.  This simulation allows us to calculate the temporal broadening effect caused by multiple reflections down the long guide channel, and this signal needs to be convoluted in time with the source pulse shape in order to calculate the systematic background shape of an ESS-like source.  This simulation is more time consuming than most others, and was run for approximately 2-4 hours on around 160 cores on the ESS computing cluster.

We now move onto the flat, photonuclear contribution.  If the guide channel and target shielding have been well designed for
background reduction, the dominant background physics could be photonuclear in nature.  Early in the design process of Target Station 2 (TS2) at the
Rutherford Appleton Laboratory (RAL, UK), it was visible in empirical studies that photonuclear channels were a major source of background, for a well-optimised target geometry, since quite simply disabling the photonuclear physics option caused the simulated backgrounds to decrease significantly \cite{ANSELL-PRIVATE-COMM}.  By empirical numerical optimisation, it was found that moving iron structures away from the moderators and replacing it with tantalum produced a marked improvement in the simulated backgrounds.  This resulted in a general idea of careful management of the material choices visible from the
back-propagated beam phase space --- particularly eliminating steels
and using high-$z$ or low-$z$ elements \cite{STUART-ICNS2013}.

Nonetheless, there was missing clarity on the physics that lead to further research.  Our colleagues at the SNS performed a study on their target geometry \cite{IVERSON-PHOTONUCLEAR} but their models found no significant photonuclear effect for prompt neutron generation.  At the ESS, a simple simulated geometry supported the idea for some time that, for prompt neutrons, there was perhaps no significant physics related to the iron nucleus but that iron may represent a ``sweet spot'' for density and total mass for objects measured in tens of cm that are found in spallation target geometries.  Both of these negative results were puzzling for a while, especially since the photo neutron background at TS2 has been measured over several hours after proton beam shutdown and is found to follow the decay of $^{56}$Mn \cite{LILLEY-PHOTO-BG-TALK-PSI}, which further supports the idea that the placement of steel around the target assembly is a strong driver of instrument background.

\begin{figure}
  \includegraphics[width=0.9\textwidth]{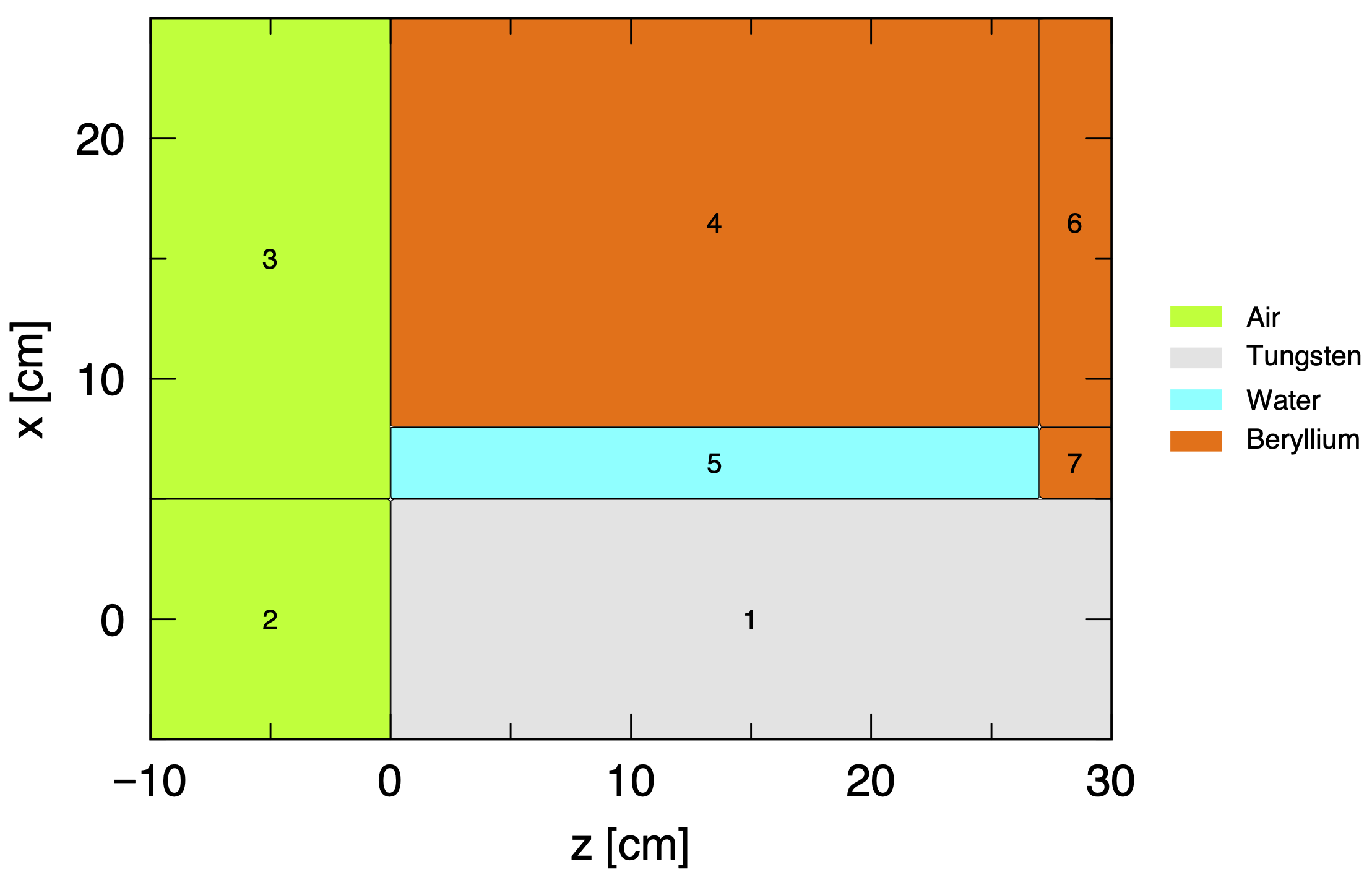}
  \caption{Simplified target geometry of an ESS-like source.  Side-view, with protons entering from the left.}
  \label{fig:photonucA}
\end{figure}

\begin{figure}
  \includegraphics[width=0.9\textwidth]{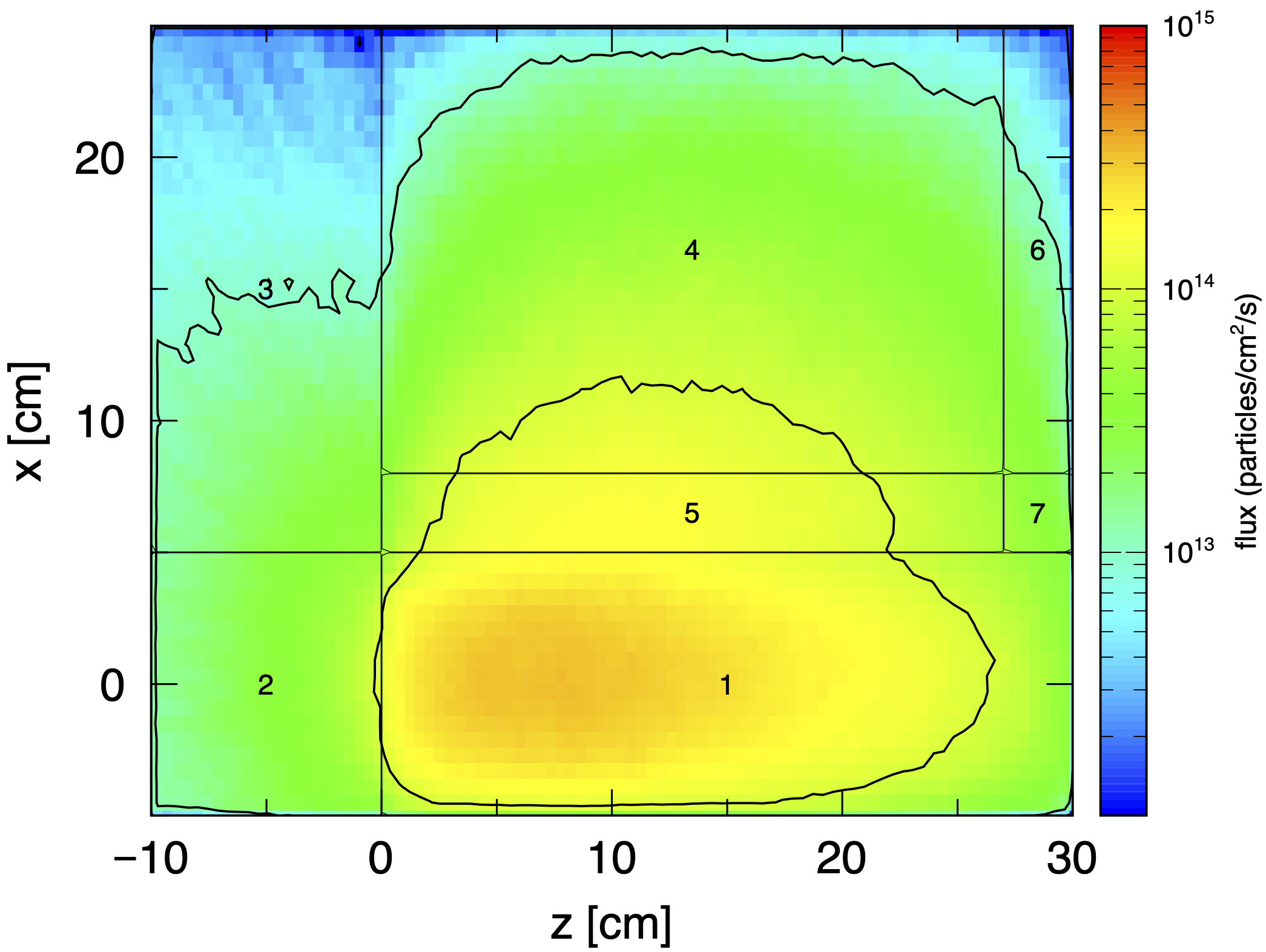}
  \caption{Neutron generation in the target as a result of proton-illumination of tungsten.}
  \label{fig:photonucB}
\end{figure}

\begin{figure}
  \includegraphics[width=0.9\textwidth]{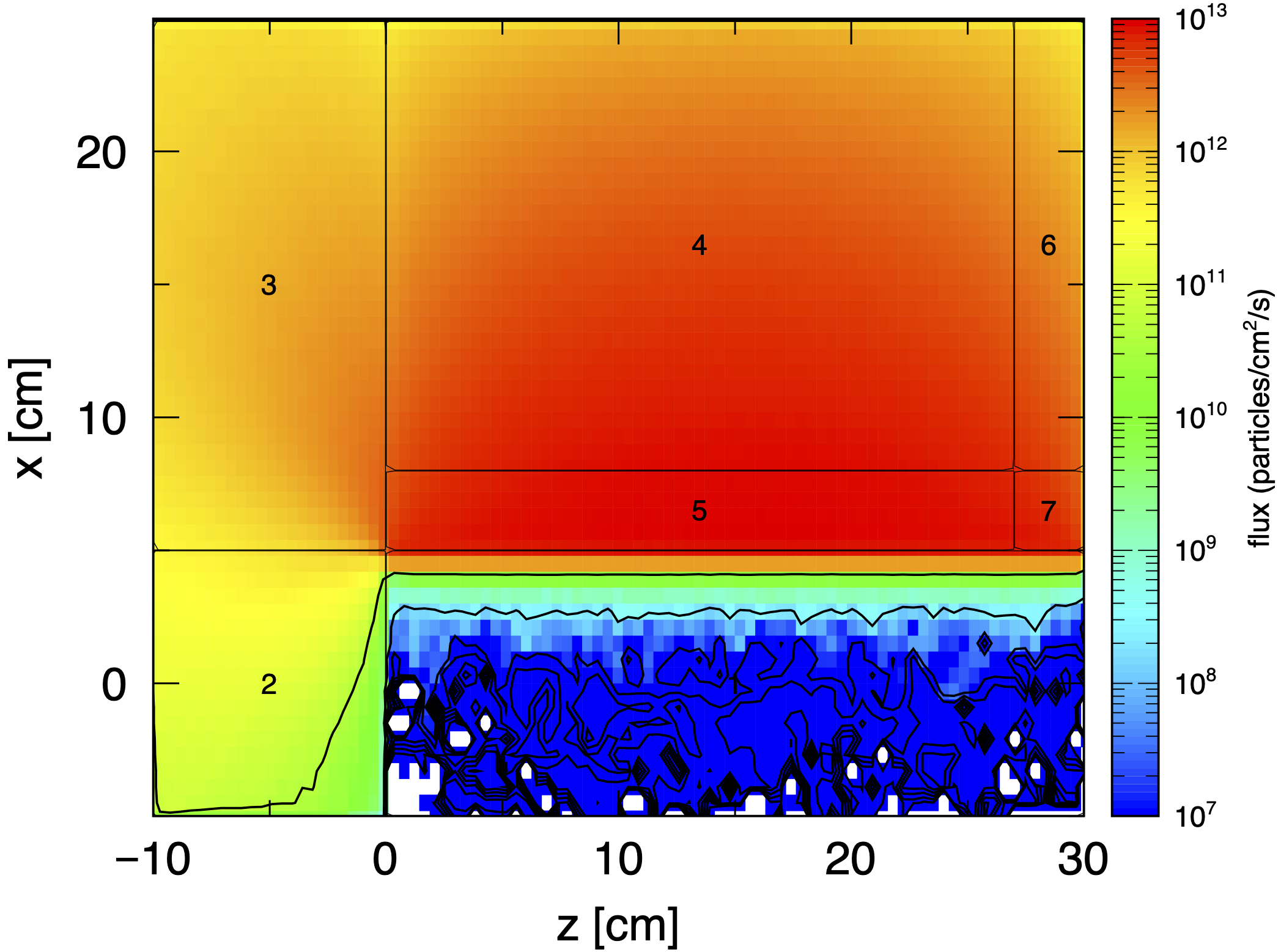}
  \caption{Gamma shine from the upper surface of the activated tungsten illuminating the beryllium reflector.  Note that the whole of the tungsten volume is activated, but only the upper surface is simulated here and biassed upwards with a $\cos{}^2$ factor to improve the efficiency of the simulation.}
  \label{fig:photonucC}
\end{figure}

\begin{figure}
  \includegraphics[width=0.9\textwidth]{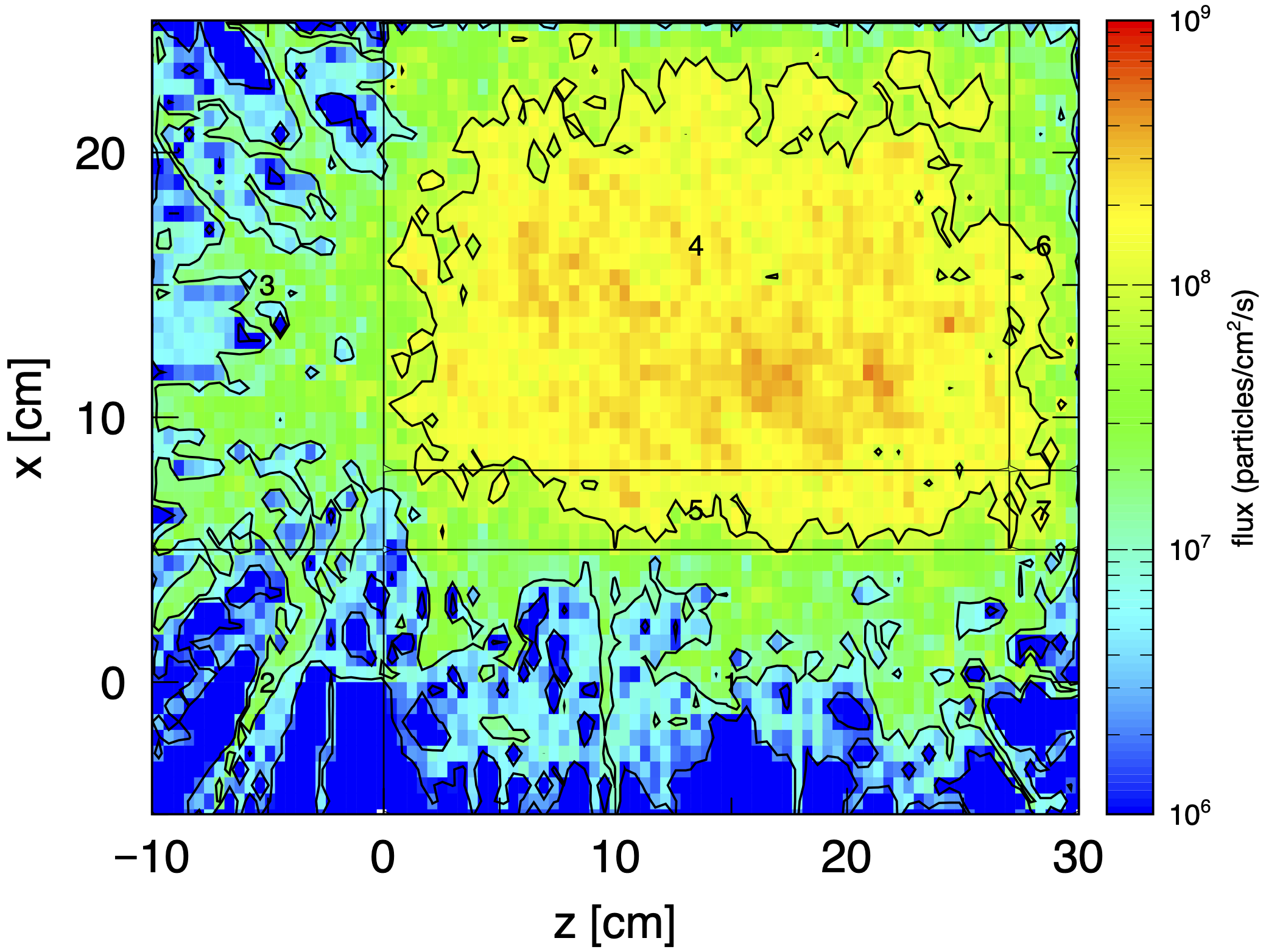}
  \caption{The resulting photoneutron production in beryllium to be compared with the prompt neutron production in figure \ref{fig:photonucB} in order to establish the flat, asymptotic fast neutron background level.}
  \label{fig:photonucD}
\end{figure}

In the present study, the photo-nuclear background component was therefore treated as a \emph{delayed} neutron source.  Establishing the flat, photo-nuclear asymptote becomes a multi-step process.  First, a simplified target geometry is created as shown in figure \ref{fig:photonucA}.  Then the fast neutron source brightness is computed from protons as in figure \ref{fig:photonucB}.  During the proton simulation, the tungsten target is used as a DCHAIN activation tally, the output of which is fed into DCHAIN at $t=0$ to generate a gamma source term for the cell.  This source is then simulated efficiently by $\cos{}^2$-biassing the uppermost 5 mm surface as shown in figure \ref{fig:photonucC}.  The photonuclear physics is activated in PHITS by setting ipnint\,=\,1, and biassed by setting pnimul\,$>$\,1.  There are problems with pnimul at values around 100, so in this simulation pnimul\,=\,50 was used.  This efficiently produces neutrons in the beryllium cells as shown in figure \ref{fig:photonucD}.  This process sounds a little involved, but was roughly a week of work in total including all debugging and researching of code options; none of the simulations took more than an hour on a workstation.

Comparing figures \ref{fig:photonucB} and \ref{fig:photonucD} one sees that for cells 5 and 7, the moderator and rear reflector surface respectively, the neutron flux in those regions is $10^8$\,$n$\,cm$^{-2}$\,s$^{-1}$ compared to $10^{13}$\,$n$\,cm$^{-2}$\,s$^{-1}$ for prompt neutrons.  This ratio is the main information that one needs, the absolute brightnesses are not important.  When taking into account the time dependence of the two components, this results in a peak maximum roughly 4 orders of magnitude above the flat component might be expected at the ESS source.

It is worth comparing this ratio with some existing knowledge.  At Los Alamos' Lujan Center, the fraction of delayed neutrons from the spallation target has been measured at 1.1$\times$10$^{-4}$ \cite{MAHURIN-SNS-DELAYED-NEUTRONS}.  At JPARC, published measurements of their background show a prompt peak standing 3.5 orders of magnitude above the flat component \emph{at the instrument}, though the precise difference there depends on proton beam configuration \cite{KIKUCHI-AMATERAS-BACKGROUNDS} and it is not immediately clear whether the dominant physics process in the flat component is indeed photonuclear in nature.  The AMATERAS instrument has a generally excellent shielding design along the beamline to reduce albedo transport, so the fast neutron backgrounds there are relatively small.

At Argonne, a uranium target was used and this would be expected to have an elevated delayed neutron channel just as it does for regular fission reactions: a delayed fraction of $\sim$3$\times 10^{-2}$ was measured \cite{ARGONNE-DELAYED-NEUTRONS}.  On LET on the second target station at RAL, the prompt peak is between 10$^4$ and 10$^5$ above the flat background.  However, in the case of that instrument, the dominant flat background effect is suspected to be ppm impurities in cadmium near the detector panels, since shielding the cadmium from the detectors produced a shadow in the background signal \cite{BEWLEY-CADMIUM-PRIVATE}.  This completely dominates all electronic noise, cosmic rays, and photo-neutrons, a result that could be expected when the photonuclear contribution was already numerically optimised to such a low level in the design of the target.  

At the ESS, the opposite optimisation to that at TS2 was necessary to shield for the compact
moderators, namely placing heavy steel shielding around the beams and moderators.  This makes the intrinsic fast neutron brightness higher because of the proximity of the target to the beamport view.  It was acknowledged that fast neutron backgrounds are a ``drawback that needs to be carefully dealt with'' \cite{ARAI-TARGET-BACKGROUND}.  One may therefore expect to see relatively elevated brightness of the ESS fast neutron backgrounds.

\begin{figure}
  \includegraphics[width=\textwidth]{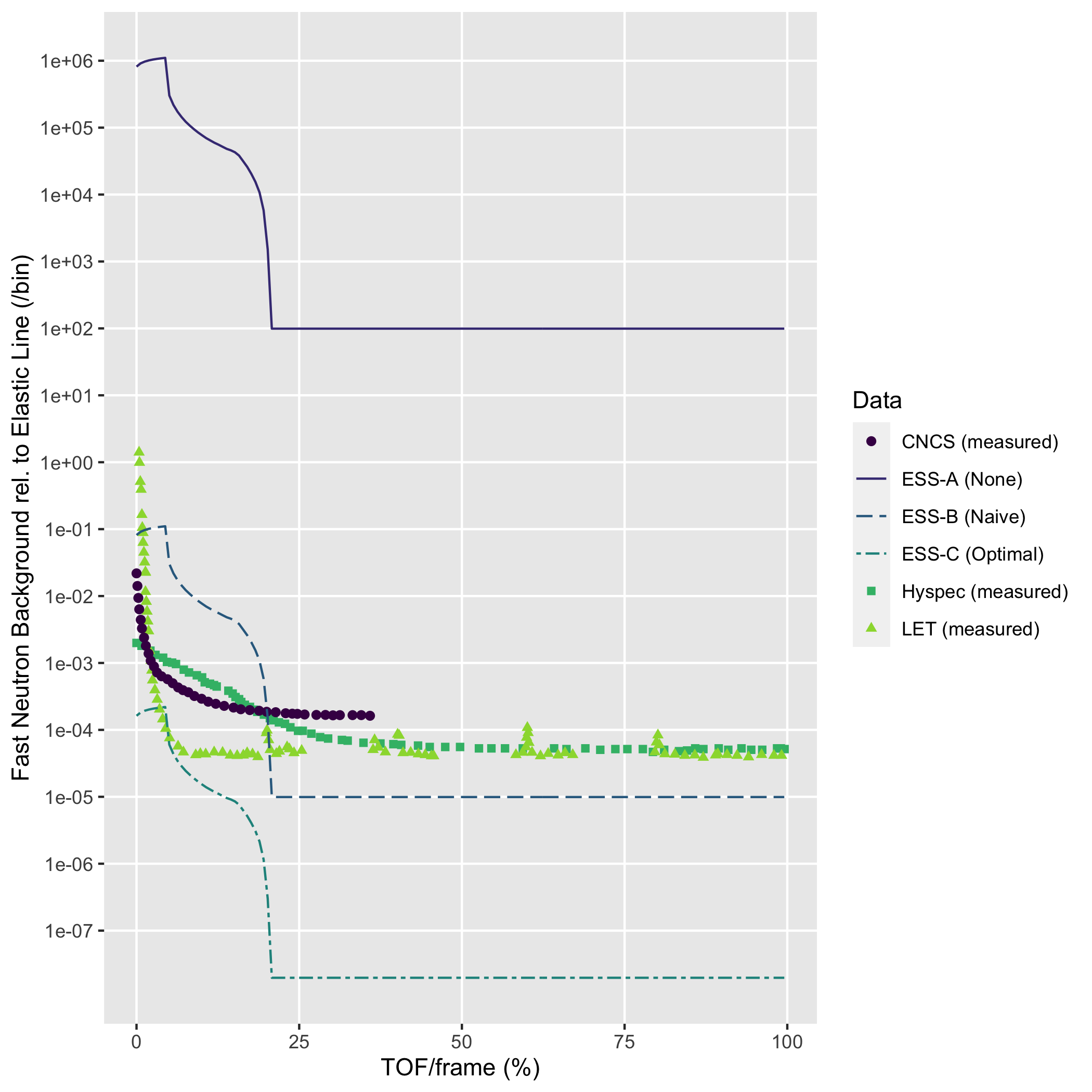}
  \caption{Comparison of existing instruments' fast neutron backgrounds with three ESS spectrometer scenarios.  The solid line ``A'' represents no design features intended to suppress backgrounds --- a straight guide --- and is the worst possible ESS case; the dashed line ``B'' represents a na\"{i}ve curved beam with no other steps taken but modelled in complete isolation from all other background effects; and the dash-dot line represents the fully optimised design for an ESS instrument taking into consideration additional technical features as described in the main article text.}
  \label{fig:essBGprediction}
\end{figure}

Using these simulated source data for ESS, one can then generate an order-of-magnitude estimate of ESS spectrometer backgrounds relative to other existing instruments.  This is shown in figure \ref{fig:essBGprediction}.  The $x$-axis is given as a percentage of the target frame to enable comparison of bandwidth and background on multiple instrument types whilst ignoring absolute temporal effects.  The $y$-axis is the size of the background count rate relative to the maximal intensity at the elastic line for a vanadium calibration.
The LET spectrometer is remarkable in that it is a straight instrument with direct line of sight to the source.  Excellent albedo reduction, using boron sheets to reduce back-scattered neutrons, and a long ``get-lost'' tube, bring the background very quickly down to the flat asymptote, and a low rep rate source provides a wide measurement window.  In comparison, the two SNS instruments HYSPEC and CNCS exhibit a lower peak height at zero time due to being curved out of line of sight, but relatively long albedo tails from the thermalising fast neutron pulse, stronger asymptotic background level, and the higher source rep-rate.

The ESS curves were estimated from multiple simulations.  The prompt curve was simulated with all neutrons generated at time t=0, and convoluted with the source pulse shape afterwards, resulting in the distorted peak at $\lesssim$ 23\% of the frame.  The flat background level was then estimated relative to the integral of that peak, as per the time-averaged relative intensities shown in figure \ref{fig:photonucA} and \ref{fig:photonucD}.  

To compare with existing instrument performance relative to vanadium spectra, the MCSTAS simulations for an ESS spectrometer were used \cite{CSPEC-SYSTEM-DESIGN}.  These indicate roughly 10$\times$ the count rate of LET.  Whilst both instrument designs exploit rep-rate multiplication (see \cite{MEZEI-REP-RATE-MULTIPLICATION} and refs therein) there is an extra normalisation factor to compare the vanadium peak height with background which is the relative number of reps in their reference configurations in the reports:  a ratio of 17/3.

Three ESS scenarios were considered.  Firstly, a worst-case, straight instrument with no attention paid to fast neutron effects in the design of the beamline, labelled ``ESS-A'' in the figure.  This obviously has a background level that is quite high compared to the peak of the elastic intensity.  No spectroscopy instruments are being built in this way at ESS, thankfully, but this curve serves to represent a useful upper limit, and it may inform the straight instruments as to their expected background level relative to the elastic measurement intensity.

Then we have ``ESS-B'' which is curved out of line of sight in a simple way, but otherwise uses no design features in shielding to attenuate fast neutrons.  The attenuation factor for this curve relative to the previous one is assumed to be 10$^{-7}$ which is known from a detailed study at JPARC of relative backgrounds for curved \emph{vs} straight guides \cite{JPARC-CURVED-GUIDES}.  Compared to other scenarios, this curve exhibits a much more intense prompt background level --- more than an order of magnitude higher than existing instruments --- but an order of magnitude lower photonuclear background in the long tails.  On average, the background is similar to existing instruments, which gives some confidence in these data.  This should, however, be viewed as a lower limit.  Firstly, generic simulations like these are idealistic and omit later engineering details that can pose significant problems.  Secondly, this treatment considers purely albedo transported neutron background, and ignores all other sources of background, which will be treated in subsequent sections. 

The final instrument shielding concept was purpose-designed for fast neutron suppression beginning in 2015, as part of a conceptual design study.  Line of sight is eliminated half way down the beamline with carefully designed collimation and beam removal features, and albedo is reduced as much as possible by a boron-rich inner surface layer of concrete \cite{MAGIC-CONCEPT-DESIGN}.  A brute-force simulation performed at PSI computed the time averaged transport of prompt neutrons.  The simulation work was performed on a dedicated cluster, with a total simulation time of two months on several hundreds of physical CPU cores.  Such a time-intensive calculation with little variance reduction is invaluable as a reliable benchmark.

The results of this simulation \cite{FILGES-IKON-TALK} lead to the expectation that there could be 4.2\,$n$\,cm$^{-2}$\,s$^{-1}$ emerging from such a system resulting from the prompt neutron physics.  This benchmark can be used to attenuate ``ESS-A'' and produce ``ESS-C'', representing a maximally efficient beam design where the integral of the peak flux within the first $\lesssim$ 23\% of the frame matches the simulation of 4.2\,$n$\,cm$^{-2}$\,s$^{-1}$.  This concept was developed as an early design exercise and was not implemented on any of the beamlines.

It is important to stress the point that these simulations quantify a purely albedo transport background effect for a single, isolated instrument.  Reducing the albedo component will require an iterative, model-based
approach to address each problem in turn:

\begin{itemize}
    \item Most importantly, fast neutron removal via a sequence of fast neutron collimators (see \cite{NOSG-HANDBOOK} page 40) --- one or two tenth values cubed of copper or steel-polyethylene laminates --- followed by large expansion zones $\sim$1\,m wide and several metres long is very effective.  This was demonstrated in a small test on the CNCS instrument at SNS \cite{OVERCOMING-BACKGROUNDS} where the addition of bags of copper shot on one side of a chopper pit resulted in a measurable 25\% reduction in background level.  More recently, the rebuild of the AMOR beamline at PSI fully exploiting these principles produced an order-of-magnitude reduction in background \cite{UWE-PRIVATE-COMM}.  The AMATERAS instrument at JPARC has used similar concepts \cite{KIKUCHI-AMATERAS-BACKGROUNDS}.
 
  \item The source photonuclear production may be minimised, by
    reducing the gamma shine on the beryllium, and moving away from
    low-volume, ``pancake'' or ``flat'' moderators and sacrificing the insignificant brightness
    gain, and instead optimising for signal to noise (noise being
    quantified as the integral fast neutron flux $\gtrsim$1\,MeV).  The
    TDR moderator design \cite{ESS-TDR}, for example, was found to have roughly the same signal to noise ratio for many instruments as the flat moderators (comparing cold neutron current and prompt, fast neutrons at the guide entrance).  This comparison is crude, and did not include a full albedo simulation of the guide system, but it was discussed amongst the experts at the time.  Indeed, the optimum beam extraction geometry aims the guide more towards the farthest edge of the moderator, as far away from the target wheel as possible, rather than the moderator center which is erroneously used in many comparisons.  In this upper position, the TDR moderator brightness is almost the same as for the flat moderator, but the shine adjacent to the target wheel is maximally reduced.  The flat moderator, meanwhile, requires aiming the guides towards a point only a few cm away from the tungsten target.   
    Ultimately, the same kind of minimisation of the backgrounds as was done for TS2 at RAL will
    need to be done for the moderator assembly, and since ESS replaces
    this unit regularly there are plenty of opportunities to do so.
    \emph{However}, a full length simulation is needed along with
    intensive numerical optimisation.  In the past, this catalysed the
    creation of bespoke tools dedicated to the task \cite{COMB-LAYER}.
  \item The ESS is expected to have a rather divergent fast neutron beam at the beamport position, caused by
    an anomalously small quantity of target shielding compared to other
    spallation sources \cite{COPPER-GUIDES}.  Extending the radius of the target monolith,
    and reversing the peculiar idea to replace steel with concrete in
    the lower half of the monolith, will reduce the target background contribution.
  \item Whilst ESS cannot use plastics near the beam port due to short
    lifetime at the elevated levels of radiation that are present
    \cite{STUART-RUBBER-LIFETIME}, TS-2 at RAL demonstrates the
    effectiveness of doing this.  There are other means to scrape the
    MeV neutron halo from the edges of the beam that are rad-hard,
    such as copper \cite{COPPER-GUIDES}.  Replacing some steel, and making more extensive use of copper
    in the beam port assemblies, would help.
    
    \item Adding a T0 chopper will reduce the prompt peak affecting the first $\lesssim$ 23\% of the frame but will not affect the relative intensity of the flat photonuclear background.  However, that may be more easily dealt with through subtraction or longer counting times depending on the magnitude of the tails: this will only be established when hot commissioning has begun.
\end{itemize}

Albedo is likely to be a strong part of the fast neutron background, and taking such steps to reduce it will have a large effect on both the prompt fast neutrons, including the pulse and the long tail behaviour, as well as the time-independent asymptote of the photonuclear background level.

\subsection{Direct Illumination \label{sec:directIllumination}}

Direct illumination of the instrument occurs when the source of fast
neutrons shines directly onto the instrument experimental area.  For
example, a target monolith building and/or bunker wall are potential
fast neutron sources that radiate over large areas into the
experimental hall.  Design of the shielding purely for safety
--- using only steel and concrete, or even worse: heavy concrete based
on iron ore --- leads to the main direct illumination
source being determined by neutron resonances in the keV-MeV energy
range \cite{IAEA-XSECT-ATLAS}.  A comparison of this specific problem
is pertinent, since the bunker was redesigned several times and an optimal, laminate-based concept was replaced with an over-simplified heavy concrete structure with the same total mass.

Concretes are not a panacea, but are often deployed as such.  They do indeed have useful and convenient roles in neutron shielding due to the ease of handling and construction, but there are established performance drawbacks.  In the past, there have been efforts to increase the hydrogen-
\cite{SERPENTINE-CONCRETE,SERPENTINE-CONCRETE2} and boron-content
\cite{COLEMENITE-CONCRETE,COLEMENITE-CONCRETE2} of
concrete.  Some existing facilities have their own borated concrete recipes.  There were some intellectual property restrictions attached to these, so an
open-source recipe was developed at ESS specifically for the purpose
of reducing fast neutron transmission \cite{CARSTEN-CONCRETE} which
contains both additional hydrogenous and borated ingredients, and it is free for people to use without commercial restriction.  It is a
1:1 replacement for regular density concrete, but requires re-bar for
structural use because there is a small reduction in strength.  At
first that might appear like a drawback, but on the scales of ESS
shielding re-bar was considered necessary for regular concrete in any case so this
is in fact option neutral.

The new material was fire-tested \cite{BPE-FIRE-TEST,BPE-FIRE-TEST2}, subjected to mechanical anchor failure testing \cite{BPE-PULL-TEST,BPE-PULL-TEST2}, and scrutinised with a range of structural and fabrication tests including a full-scale industrial production \cite{DTI-CONCRETE-REPORT}.  These extensive tests were performed in part associated with the development but also in response to some technical risk concerns that were raised, and with all the risks thus addressed in the referenced reports the material was demonstrated to be viable.

There were two baseline laminate designs considered by the expert
scientific team for bunker shielding at ESS.  Both of them had similar
neutronics performance.  The first was a steel-polyethylene laminate;
the second replaced the polyethylene with the BPE-concrete mix described above to
some of the practical advantages of dealing with concrete rather than
polyethylene.

A subsequent bunker design initiative by Mezei \cite{MEZEI-BUNKER-DESIGN} replaced these laminates with a single volume of high-density concrete, assembled from blocks of identical composition, using a mix under the commercial name ``magnadense''.  The marketing materials for this material claimed a density of up to 3900\,kg/m$^3$ whilst the density ultimately covered by the contract is 3800\,kg/m$^3$.

\begin{figure}
  \includegraphics[width=0.9\textwidth]{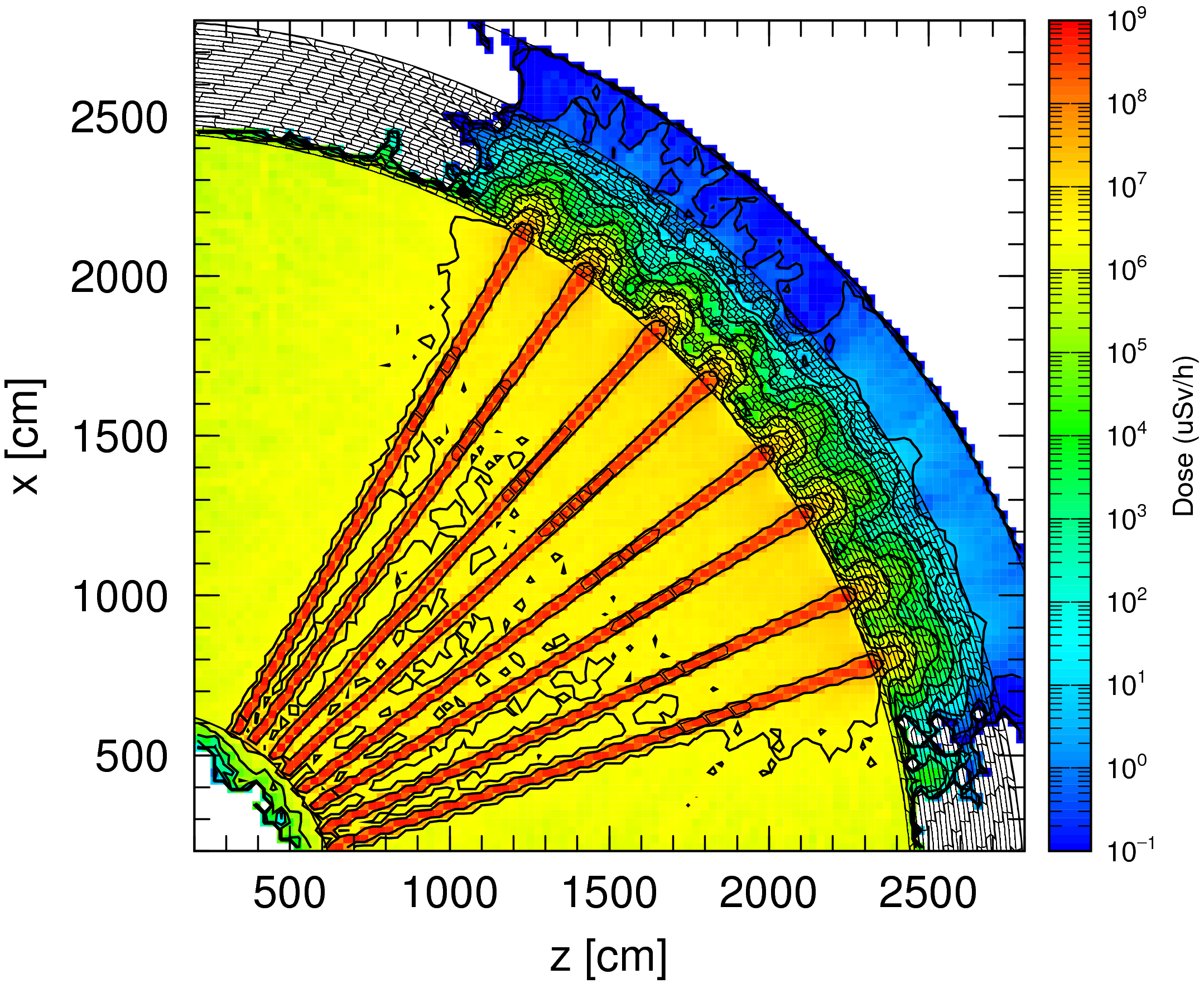}
  \caption{PHITS simulation of the baseline, laminate bunker design illuminated
    by a full sector of eight instruments.  The top four beamlines
    have heavy steel inserts in the bunker wall, and the lower four
    beamlines have the inserts removed.}
  \label{fig:OriginalBunker}
\end{figure}

\begin{figure}
  \includegraphics[width=0.9\textwidth]{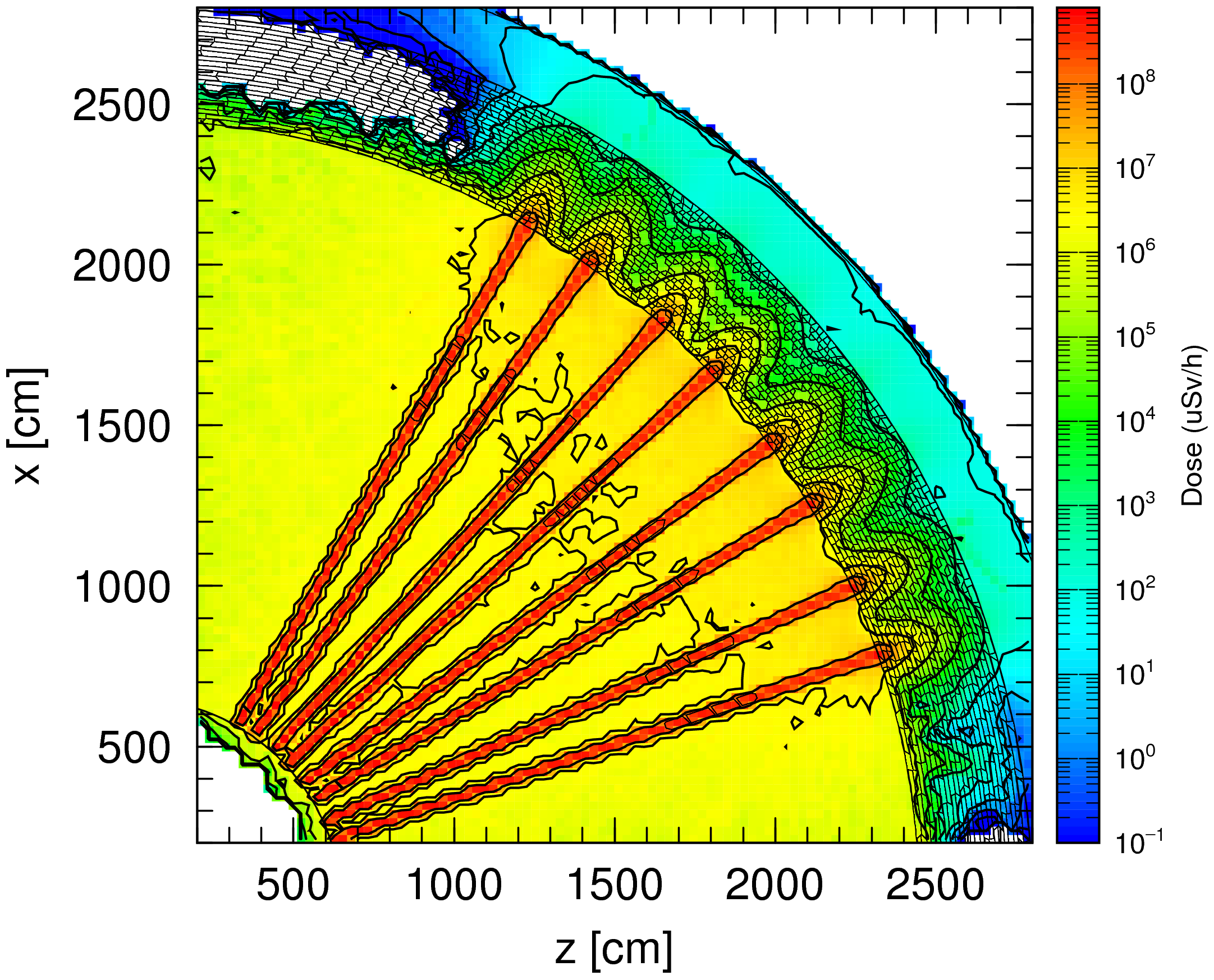}
  \caption{PHITS simulation of a heavy concrete bunker wall
    illuminated by a full sector of 8 instruments.}
  \label{fig:HCbunker}
\end{figure}

To explain why heavy concrete is a poor choice, figure \ref{fig:OriginalBunker} shows the baseline bunker design with steel and a BPE concrete laminate in two configurations, one with
large steel inserts in the wall (upper four beam ports in the figure), and the other without steel inserts (lower four beam ports in the figure).  The lower four beamports are those one should focus on for the comparison, since the steel inserts were removed from the considered options by the project.  Here the external dose rate is $\sim$10\,$\mu$Sv/h taking no safety
credit for any equipment in the beam --- this is the correct procedure to estimate the wall performance
from a nuclear safety perspective: unless equipment is interlocked one
cannot assume it is present, otherwise there can be unsafe scenarios
during repair work when equipment is taken away.  This important point will be addressed in more detail shortly.
This should be compared with the data in figure
\ref{fig:HCbunker}, which shows the expected changes resulting from the use of heavy concrete with the same total mass.

\begin{figure}
  \includegraphics[width=0.9\textwidth]{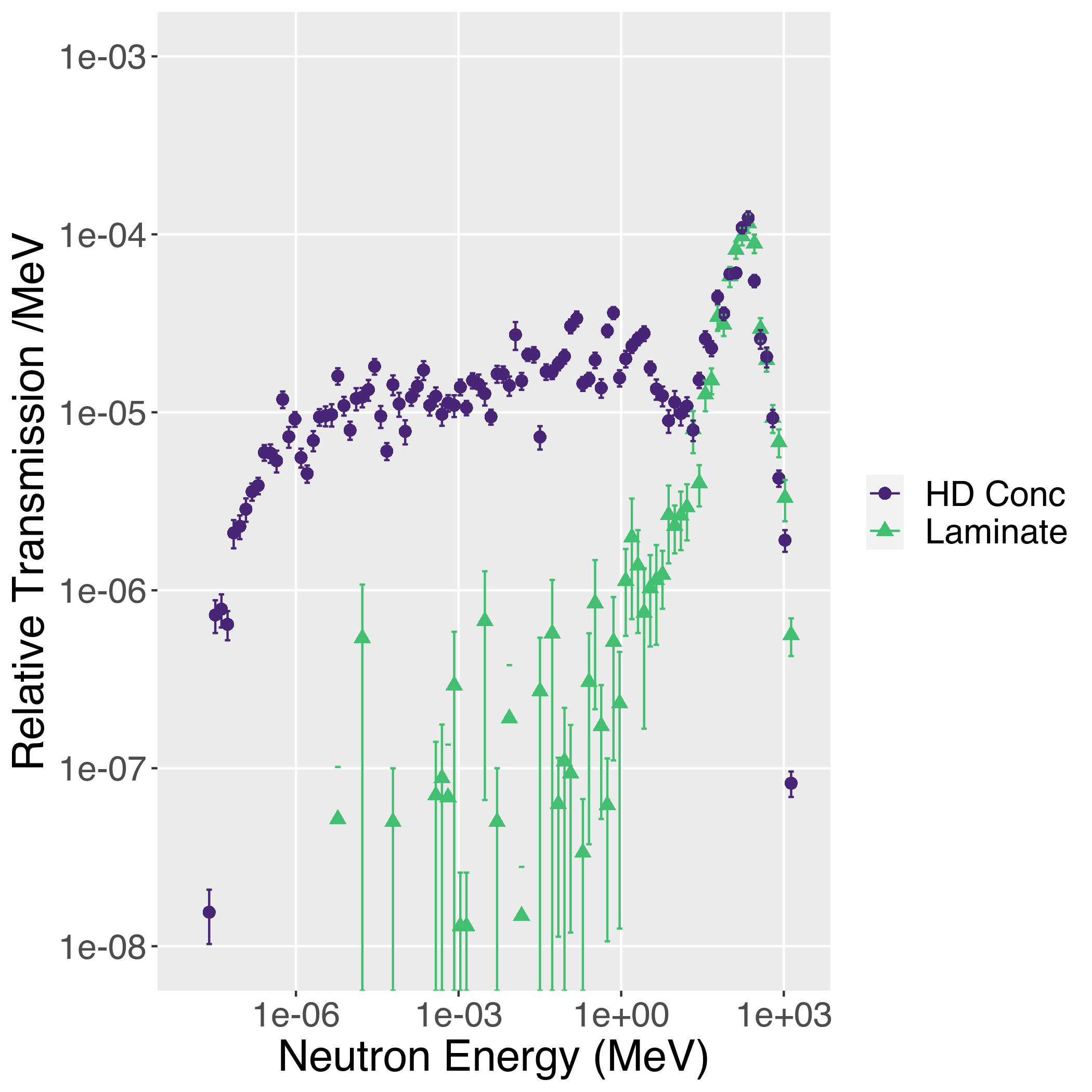}
  \caption{PHITS simulation of transmitted neutron spectrum relative
    to the input spectrum, through a heavy concrete bunker wall (``HD Conc'')
    \emph{vs} a simplified version of the original baseline design of
    a laminate of steel and borated polyethylene (``Laminate'') as indicated in the
    legend.}
  \label{fig:whyNotHeavyConcrete}
\end{figure}

Those who read the original heavy concrete proposal \cite{MEZEI-BUNKER-DESIGN} may be confused initially, since report describes a significantly reduced total mass and initial cost, which makes the option attractive.  The dose rate external to shielding illuminated by neutrons approaching and exceeding the Hagedorn temperature is dependent, mainly, on the line integral of the shielding mass.  It was therefore inevitable that the shielding mass would gradually increase to the other existing options, to achieve the safety requirements.  This is visible in the comparison of the spectral components emerging from both the laminate and heavy concrete wall, shown in figure
\ref{fig:whyNotHeavyConcrete}.  The overlapping peak at high energy corresponds to neutrons around the Hagedorn temperature.  The additional neutron flux affecting backgrounds is where the two spectra diverge, below 1\,MeV, as one would expect.

We can now quantify the direct shine component to those instruments that are potentially illuminated.  Our fast neutron measurements at PSI \cite{PSI-TARGET-NEUTRONICS} revealed a dose rate of around 0.04~$\mu$Sv/h from their target shielding monolith wall
measuring 14$\times$9.6 metres ($a_1$) corresponding to 280 \nsms.  It
is fairly trivial to calculate that the instrument flux ($\approx a_1
a_2 / r^2$) for a 5$\times$5~m$^2$ ($a_2$) unshielded instrument
$r=30$~m away, illuminated by this PSI target monolith directly, would
be above $10^2$~\nps fast neutron background current.  A more accurate
cylindrical correction could be used for this calculation, as well as
groundshine albedo corrections, but for an order of magnitude estimate these are not necessary.  We can conclude from these numbers, in comparison to those discussed for the SNS instruments earlier, that PSI benefits from the fact that it is a continuous source, and so the fast neutron
background is entirely a statistical error rather than a systematic error.  PSI also
has more substantial shielding structures around some of its
instruments in comparison to HYSPEC and CNCS, and together these allow PSI to operate with relatively high fast neutron background levels (that are still well below radiologically safe limits) in comparison to pulsed sources.

\begin{figure}
  \includegraphics[width=0.9\textwidth]{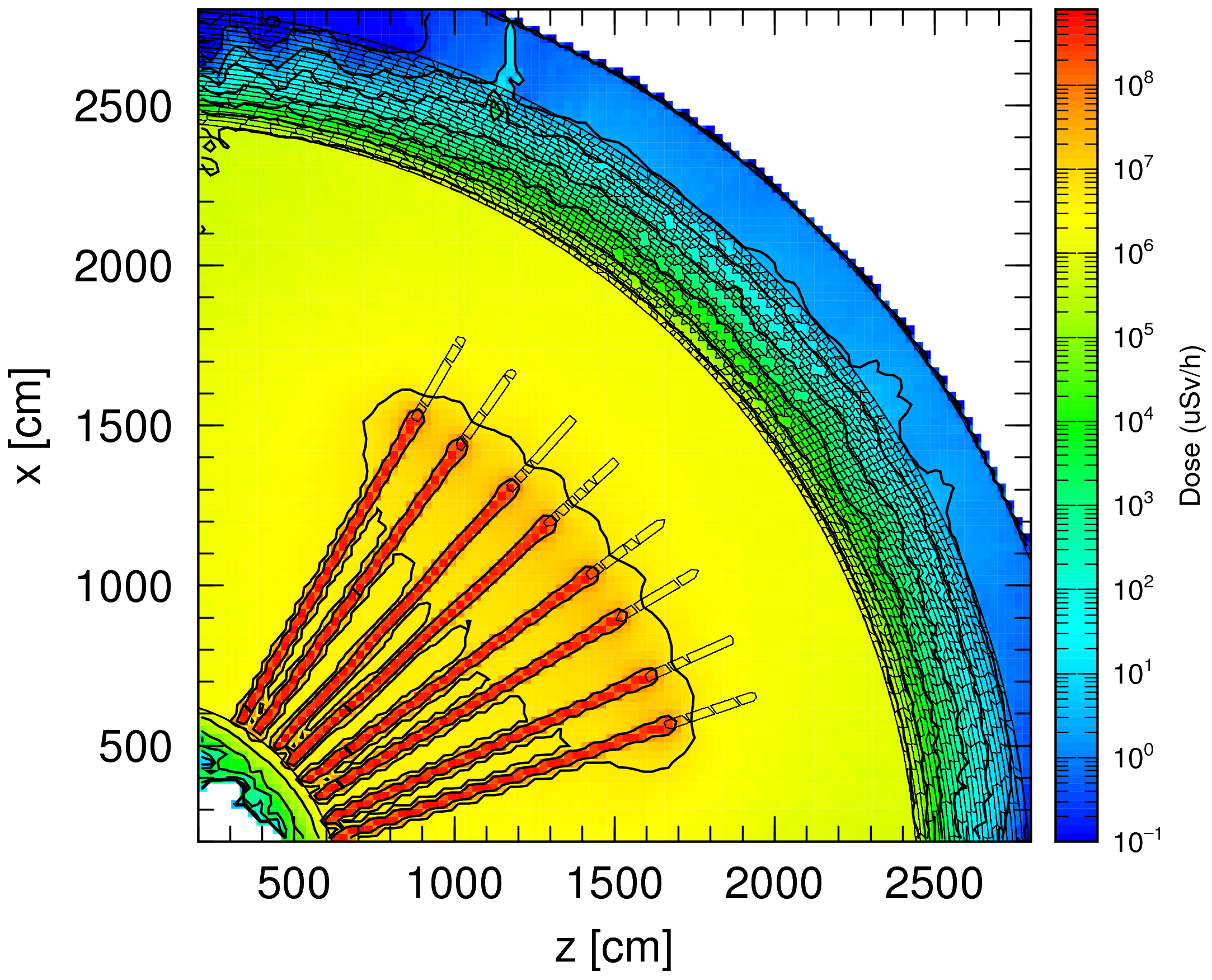}
  \caption{PHITS simulation of the heavy concrete ESS bunker with the entire beam
    blocked by instrument hardware.  The external dose rate at the wall surface is 
    10\musvh ($\implies \sim 10^5$\nsms.)}
  \label{fig:bunkerShuttersClosed}
\end{figure}

As mentioned before, the simulations of the ESS bunker indicated in figure \ref{fig:HCbunker}, are purely for safety assessment of the bunker wall thickness.  Under those circumstances,
one must not take safety credit for any hardware that is not part of
the safety validated set of components, necessitating interlocks or
other safety guarantees.  As such, the bunker void is completely empty.  The model
also removes the wall penetrations that normally allow the beam to
pass through to the instrument, so that the shielding properties of
the wall are assessed in isolation of other effects.  A valid critique
of these data from a background point of view is that the real bunker
will contain hardware, which will deflect the fast neutrons to some
extent, and a penetration through the wall will reduce scatter.  The
author accepts these points --- it has always been maintained that two
sets of neutronics simulations need to be performed: one for safety,
and one for background, to take into account both limiting scenarios.
To this end, whilst still maintaining a generic viewpoint, a
hypothetical set of heavy shutters was simulated with all beams in
the closed position, and all guide penetrations through the bunker
wall are also closed.  This represents an optimistic shielding
scenario where the vast majority of fast neutrons are scattered into
the bunker void, and the results are shown in figure
\ref{fig:bunkerShuttersClosed}.

Considering that the tenth value for fast neutrons is roughly 40~cm of
steel \cite{SULLIVAN} and whilst there are plenty of long components
in the bunker exceeding this metric, the lack of hardware with a
\emph{transverse dimension} comparable to a tenth value indicates that
any scattered neutrons will spread through the bunker like a low
pressure gas, since 3-4 albedo \cite{SELPH-FAST-NEUTRON-ALBEDO} reflections are required to reduce the intensity by an order of magnitude.  This is indeed what one
can clearly see in figure \ref{fig:bunkerShuttersClosed}, by the
uniform yellow shading throughout the bunker.  Moreover, the scattered beam inside the bunker that does not penetrate the wall does not magically disappear, instead just under half of it would be expected to contribute to the roof surface source of skyshine (more details on skyshine are given in section \ref{sec:skyshine}).

Focussing our attention still on the wall, in this maximally scattered fast neutron scenario, the external dose
rate of the bunker wall is 10~\musvh (which implies a fast neutron flux around $\sim$10$^5$\nsms).  In reality, it is likely that an uneven distribution of instrument hardware, both internal and external to the bunker wall, will cause local variations to these figures.  These give an indication of background levels to be expected near the bunker wall surface, as it was known at
the time in 2018 \cite{ESS-0321186}, with the lower figure of a fast neutron flux around $\sim$10$^5$\nsms{} seen to be the most likely value at many hotspots.  A more detailed, independent benchmark would of course be invaluable.

This was done meanwhile at PSI for the ESTIA
instrument \cite{ESTIA-TG3-ESS-0432926}.  Three important points must be
borne in mind: only a single activated beam port was modelled; the short sector for ESTIA has a wall radius around 15 metres rather than 25 metres as shown in previous sections; and the team have worked hard to ensure that
the instrument avoids direct line of sight of the target source
through a sequence of well-shielded zig-zag optical components.  To
compare with the previous two data sets, one must look at the top most
beam port in figures \ref{fig:HCbunker} and
\ref{fig:bunkerShuttersClosed}.  The PSI data indicate a minimum dose
rate of fast neutrons around 0.1~\musvh{} ($\approx 10^3$ \nsms) into
the experimental cave at the farthest point away from the beam axis,
rising to 10~\musvh{} within the fast neutron halo around the wall
penetration ($\approx 10^5$~\nsms).

The former number is lower than those for the generic cases, as one
would expect, and the latter is in excellent agreement with the
maximally-scattered scenario from the 2018 study, shown in figure
\ref{fig:bunkerShuttersClosed}.   Before applying this number to other beamline estimates, one must
consider the area of the 10$^5$\nsms{} source.  For an order of
magnitude estimate, 1$\times$1\,m$^2$ seems to be at the low end of
the range, and 3$\times$3\,m$^2$ for a SANS or reflectometer front
wall are at the upper end.  These two areas span only one order of
magnitude nonetheless, so an unmodified bunker wall fast neutron
background current of $\sim$10$^5$~\nps{} is a good starting point,
which is the result of two independent simulations using different
simulation software.

Reducing this component of the background would require replacing, or
augmenting, most of the shielding material around the target / bunker
region and restoring some of the laminate concepts from 2017 that
specifically target fast neutrons as being a key problem in shielding
design.  This might not be as expensive as it sounds at first.  Once
again one can look at TS-2 at RAL, and see the liberal use of borated
polyethylene and paraffin wax on some external shielding surfaces tens
of cm thick.  Of course, such steps would require fire risk
management, these carry additional costs that SNS chose to avoid.
With this knowledge, very early in the ESS project we engaged with ESS fire safety experts who investigated \cite{RAL-WAX-CAN-KNOWLEDGE-TRIP} and fire tested \cite{PARAFFIN-CAN-FIRE-TEST} steel
cans filled with paraffin wax.

RAL builds caves almost entirely from such wax tanks, since the gamma output of the instruments is low compared to ESS and the highest energy neutrons are in a very short pulse.  For ESS, with a large time average flux and a long pulse --- which create a large gamma load at most sample positions, and a GeV energy neutron pulse that is milliseconds in length --- the wax can concept for instrument caves is probably only viable when deployed in the form of laminate solutions with steel.  Some version of this concept may be used for new instrument projects down the line.

\subsection{Skyshine\label{sec:skyshine}}

Skyshine is essentially fast neutron albedo from the atmosphere, the
physics is the same but the density of the scattering medium is much
lower.  As such, the particle range is not measured in cm or metres
but rather several hundred metres.  Stevenson and Thomas
\cite{STEVENSON-SKYSHINE} reviewed a range of studies of particle
accelerators in the 1980s, and observed that the fast neutron flux
escaping into the atmosphere was well described by the following
equation:
\begin{equation}\label{eq:skyshine}
  \phi(r) = \frac{\Phi A}{4\pi r^2} e^{-r/\lambda}
\end{equation}
where $\Phi$ is the source strength (in units of \nsms) and $A$ the
effective area of the source; $r$ is the distance from the source (m);
and $\lambda$ is the effective absorption length (m) that depends on the neutron energy.  $\lambda$ is
observed to be in the range $\lambda=300-900$~m for spallation source
energies.  This function is useful for ESS since the linac is several
hundred metres long and the long beamlines are $>$150~m in length.

\begin{figure}
  \includegraphics[width=0.9\textwidth]{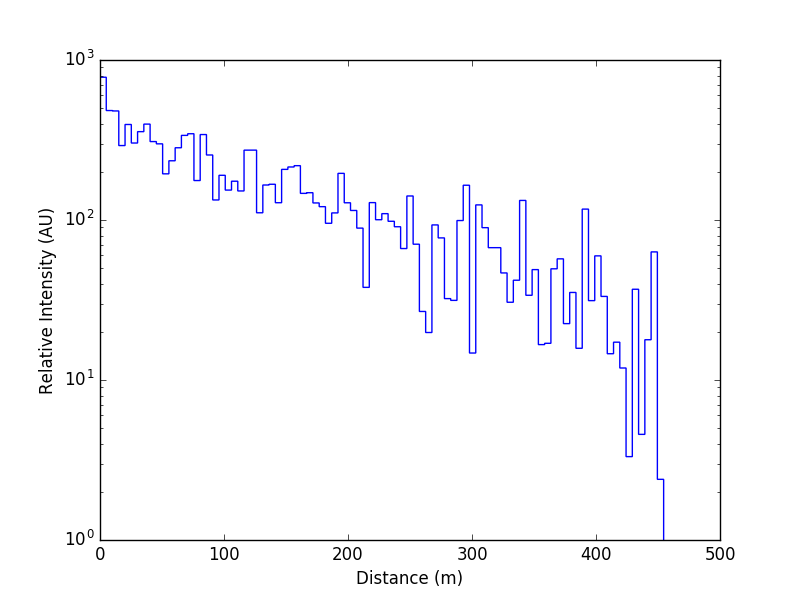} \\
  \includegraphics[width=0.9\textwidth]{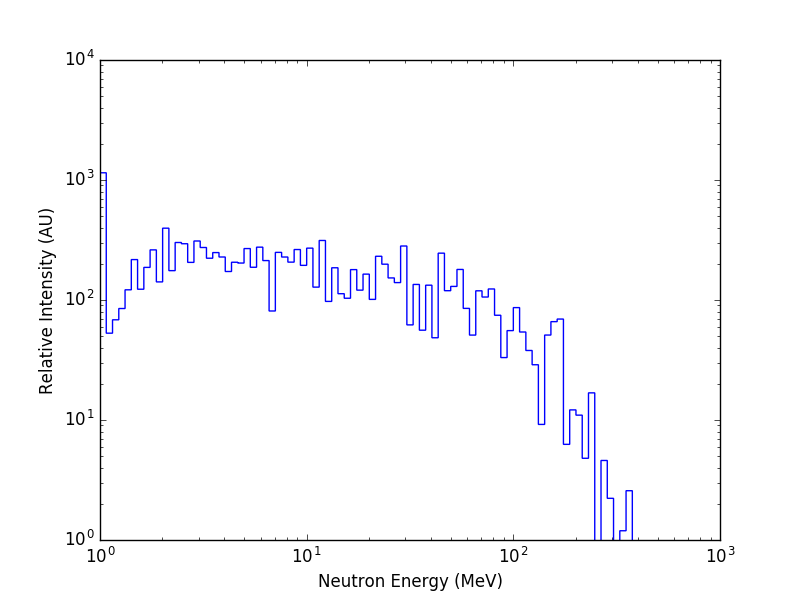}
  \caption{GEANT4 simulation of skyshine from the ESS, using a single LINAC
    stub of the accelerator with the highest neutron energy. (TOP): the variation with distance
    $r$ of neutrons of any energy; (BOTTOM) the variation with energy
    of neutrons at any distance.}
  \label{fig:GEANT4-SKYSHINE}
\end{figure}

Skyshine is known to be a radiological issue at particle accelerator facilities, and so is part of the safety assessments with nuclear authorities.  One can therefore use documented source terms from the safety reports to construct a theoretical and/or simulated skyshine radiation field and assess its potential impact on instrument backgrounds.  
Two points were raised on this topic during the design phase of ESS.  Firstly, the distances involved --- hundreds of metres --- are significantly greater than the mean free path of thermal neutrons in air, and this can lead to an underestimation of the potential significance of the mechanism.  A second problem was a misunderstanding that the returning neutrons are only of low energy and can therefore be ignored --- in fact they cover a range of high energies.  To answer both of these points, therefore, ESS skyshine was simulated using an early source term from the accelerator, using GEANT4
\cite{GEANT4} as shown in figure \ref{fig:GEANT4-SKYSHINE}.
The figure shows clearly that the returning neutrons cover the whole energy spectrum, and that they shine onto the entire spallation source site.

In subsequent sections, more updated source data are used as supplied in public licensing documents, but the methodology is the same, and these are now released for the target \cite{TARGET-SOURCE-TERM}, bunker \cite{BUNKER-SOURCE-TERM} and accelerator \cite{ACCELERATOR-SOURCE-TERM}.

\subsection{Hotspots}

In addition to these large structure sources, there are to be expected a number of small, yet strong, sources of fast neutrons at various locations around the facility.  They could be cable and piping openings, design changes during construction, weak points in the construction etc.  One such prominent location was identified in a survey at the SNS (\cite{SNS-SURVEY}, figure 1) as being a gap between the main target shielding and the external shielding above the beamlines, which is to allow the structures to move independently in the case of an earthquake.  It was identified as an MeV source shining on a structure above one of the instruments reporting background problems.  Simulations of this earthquake gap for ESS  \cite{ESBEN-EARTHQUAKE-GAP} are broadly in agreement with the measured dose rates at the SNS \cite{SNS-SURVEY}, in the range 10--30\,$\mu$Sv\,hr$^{-1}$.  That the majority of the neutrons are in the $\geq$\,MeV energy range was experimentally verified at the SNS by the use of multiple neutron detector types with different energy-dependent response functions.  The ESS uses the same concept as the SNS, despite these known background impacts, so we can quantify this contribution and also use it to compare with SNS.

\subsection{Main Background Results\label{sec:ESSBackgroundLevels}}

We can now consider possible background contributions for a source like ESS, calculated from the work in all previous sections, and using as sources the various contributions from each of the facility components as estimated in internal safety reports as cited.  This represents a numerically estimated breakdown of the ``everything else'' bar from the SNS that was shown in figure \ref{fig:SNS-PARETO}.

\begin{figure}
  \includegraphics[width=0.9\textwidth]{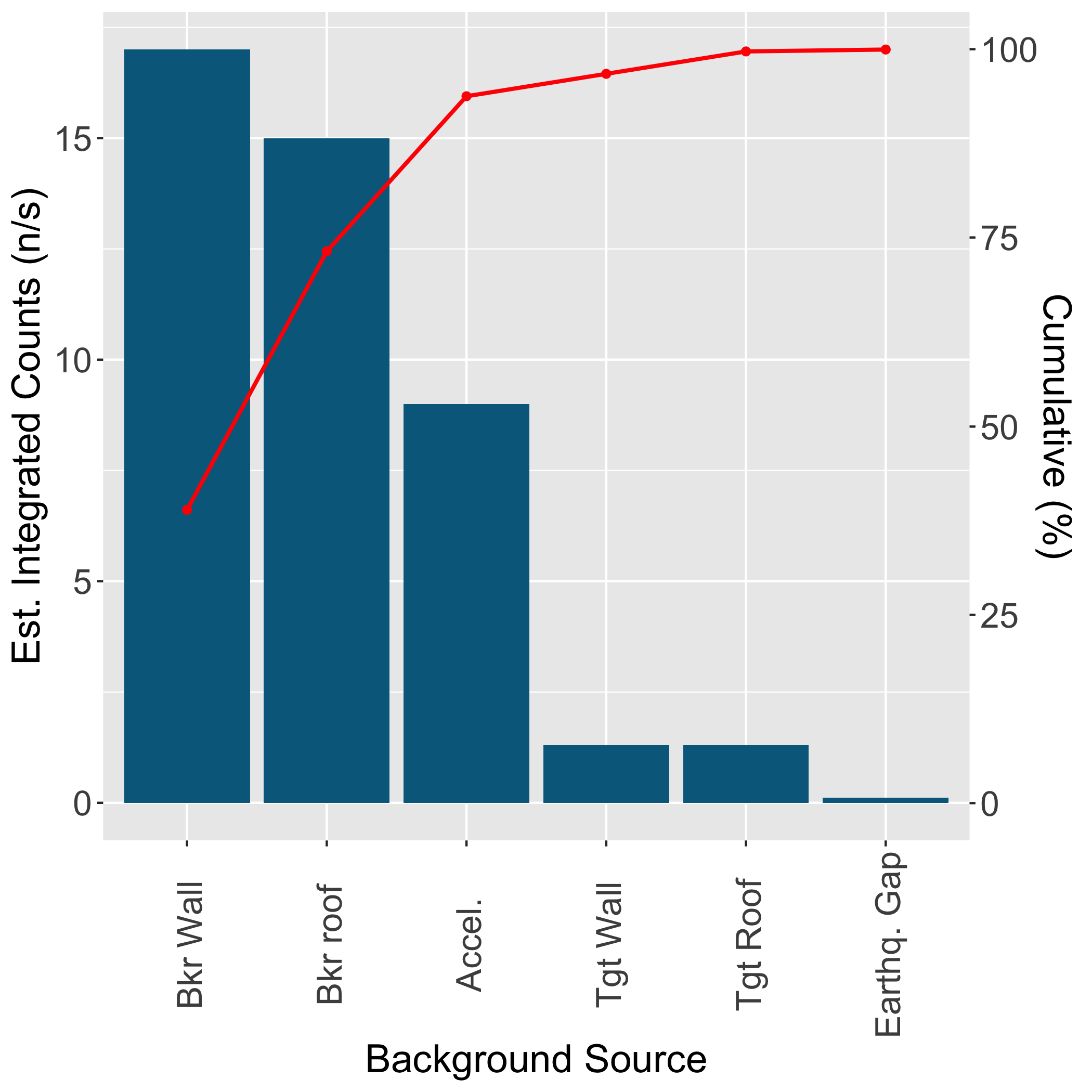} \\
  \includegraphics[width=0.9\textwidth]{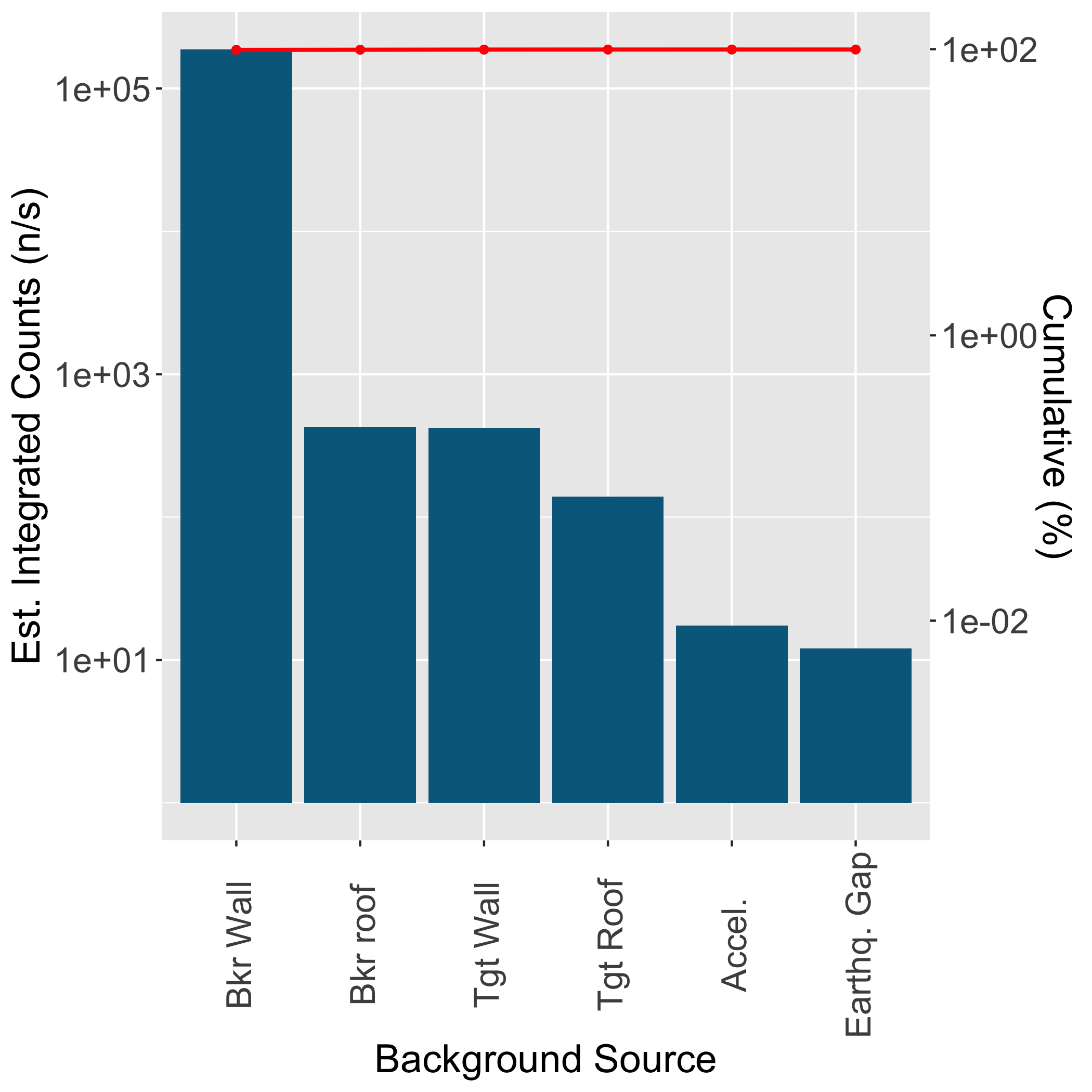}
  \caption{Pareto charts of expected background contributions to generic long ESS instruments (top) and short ESS instruments (bottom) as described in the text.  The lower graph is on a log-scale on the $y$-axis since the bunker wall strongly dominates.}
  \label{fig:ESS-PARETO}
\end{figure}

The external contributions assume a cave-detector response factor of 0.1 (from section \ref{sec:caveResponses}).  It is also assumed that short instruments do not build their cave directly opening from the outer surface of the bunker wall but instead add their own inner wall of some kind, with the same 0.1 response factor.  This is shown in figure \ref{fig:ESS-PARETO}.  Any instruments that do not add additional shielding downstream of the bunker will experience the full background flux that the bunker wall may offer in some locations.

Note that instrument cross-talk is ignored for simplicity, but there may be hotspots on neighbouring beamlines that are significant.  At the SNS, two other beamlines were contributing $\gtrsim$10\% in figure \ref{fig:SNS-PARETO}, and these would also need to be reduced by an order of magnitude or more in a successful background reduction campaign.

What figure \ref{fig:ESS-PARETO} tells us is that for long instruments, after the albedo background of their own beamline, the most likely sources are the bunker and accelerator.  On the other hand, for the short instruments, the bunker wall is the most significant external source of background.  The earthquake gap at the ESS is the smallest expected contributor in the simulations, despite being similar in strength as a source on an absolute scale.  This indicates that backgrounds generally are expected to be higher than those seen elsewhere.

On a final point, as mentioned ealier in the paper, alternative neutron detectors to $^3$He might behave differently in terms of high energy background.  $^{10}$B detectors, for example, offer an order of magnitude fast neutron background reduction \cite{BORON-DETECTOR-BACKGROUND}.  This is a very rough comparison of course, and more detailed simulations are needed overall, but as part of a package of work this may be an excellent option.  For example, if one is searching for two orders of magnitude background reduction, one order of magnitude might be achieved with a piece of steel $\sim$40\,cm thick \cite{SULLIVAN} in the right place (or indeed removing the same amount of material in the wrong place!), and switching to $^{10}$B detectors.  That could be cheaper overall than trying to find two orders of magnitude with shielding alone, by deploying $\sim$80\,cm and having to manage significant floor loading changes etc.

\section{Conclusions}

A broad range of shielding simulations have been produced to estimate various contributions to instrument backgrounds, to anticipate shielding design issues and challenges facing the European Spallation Source project in early operations.  It is found that, in a simplified, na\"{i}ve scenario, which could nonetheless be considered as a best-case scenario from the deployed concepts, the signal to noise ratio of instruments that are curved out of line of sight is broadly similar to instruments at other facilities that report problematic background levels, except during and just after proton illumination, where the ESS may see backgrounds that are elevated by an additional 1-2 orders of magnitude.  Straight instruments appear to suffer the most, as one would expect, with background levels varying between 2 and 6 times greater than the elastic signals.  With the presence of non-idealistic, real-world geometries, some additional increases could be expected than those simulated here.

For each major physical contribution mechanism, a list of potential avenues to explore in order to reduce the backgrounds was given, and a breakdown of the relative strengths of these contributions was also provided.  For both the long and short instruments, the bunker has become the dominant external source of background, after albedo transport down the instrument's own guide system.  It may be necessary to replace this structure with a laminate design based on spallation physics, rather than a reactor-style design that currently exists.

%
%
%

\section{Acknowledgements}
The author is grateful to a number of collaborators, for useful discussions
and for establishing the groundwork upon which this paper could rest.  S. Ansell, U. Filges, K. Kanaki, R. Hall-Wilton,
C. Zendler, V. Santoro, N. Cherkashyna, D. DiJulio, E. Rantsiou, 
C. Cooper-Jensen, D. Martin Rodriguez, R. Kolevatov, G. Greene, J. Cook, D. Mildner, E. Iverson, R. Bewley, K. Fissum, C. Dewhurst, and S. Lilley are thanked in particular.

\bibliographystyle{unsrt} 
\bibliography{earlyOps}

\begin{thebibliography}{10}

\bibitem{ESS-TDR}
S.~Peggs (Editor).
\newblock {ESS} technical design report.
\newblock Technical report, Lund University ESS-doc-274-v15 / ISBN
  978-91-980173-2-8, 2013.

\bibitem{NOSG-HANDBOOK}
C.~Zendler, V.~Santoro, S.~Ansell, N.~Cherkashyna, D.~Martin Rodriguez,
  C.~Cooper-Jensen, D.~DiJulio, and P.~M. Bentley.
\newblock European spallation source neutron optics and shielding guidelines,
  requirements and standards.
\newblock Technical report, {ESS-0039408}, 2015.

\bibitem{BENTLEY-COST-INTERNAL-REPORT}
P.~M. Bentley.
\newblock An initial cost-benefit analysis for instrument optics and shielding.
\newblock Technical report, {ESS-0039333}, 2015.

\bibitem{BENTLEY-COST-OPTIMISATION}
P.~M. Bentley.
\newblock Instrument suite cost optimisation in a science megaproject.
\newblock {\em J. Phys. Commun.}, 4:045014, 2020.

\bibitem{LUE-200}
A.~Sumbaev, V.~Kobets, V.~Shvetsov, N.~Dikansky, and P.~Logatchov.
\newblock {LUE-200} accelerator --- a photo-neutron generator for the pulsed
  neutron source ``{IREN}''.
\newblock {\em Journal of Instrumentation}, 15:T11006, 2020.

\bibitem{SNS-SPECTROMETERS}
M.~B. Stone, J.~L. Niedziela, D.~L. Abernathy, L.~DeBeer-Schmitt, G.~Ehlers,
  O.~Garlea, G.~E. Ganroth, M.~Graves-Brook, A.~I. Kolesnikov, A.~Podlesnyak,
  and B.~Winn.
\newblock A comparison of four direct geometry time-of-flight spectrometers at
  the spallation neutron source.
\newblock {\em Review of Scientific Instruments}, 85:045113, 2014.

\bibitem{CNCS}
G.~Ehlers, A.~A. Podlesnyak, J.~L. Niedziela, E.~B. Iverson, and P.~E. Sokol.
\newblock The new cold neutron chopper spectrometer at the spallation neutron
  source: Design and performance.
\newblock {\em Review of Scientific Instruments}, 82(8):085108, 2011.

\bibitem{BEWLEY_LET}
R.~I. Bewley, J.~W. Taylor, and S.~M. Bennington.
\newblock {LET}, a cold neutron multi-disk chopper spectrometer at {ISIS}.
\newblock {\em Nuc. Inst. Meth. Phys. Res. A}, 637:128, 2011.

\bibitem{MEZEI-LONG-PULSE}
F~Mezei.
\newblock The raison d'etre of long pulse spallation sources.
\newblock {\em J. Neutron Research}, 6:3--32, 1997.

\bibitem{COPPER-GUIDES}
P.~M. Bentley, R.~Hall-Wilton, C.~P. Cooper-Jensen, N.~Cherkashyna, K.~Kanaki,
  C.~Schanzer, M.~Schneider, and P.~B\"{o}ni.
\newblock Self-shielding copper substrate neutron supermirror guides.
\newblock {\em J. Phys. Commun}, accepted 2021.

\bibitem{SNS-BACKGROUND-SURVEY}
M.~B.~R. Smith, E.~B. Iverson, F.~X. Gallmeier, and B.~L. Winn.
\newblock Mining archived {HYSPEC} user data to analyze the prompt pulse at the
  {SNS}.
\newblock Technical report, ORNL/TM-2015/238, 2015.

\bibitem{MICHEL-ACTIVATION}
R.~Michel, D.~Hansmann, S.~Neumann, W.~Glasser, H.~Schuhmacher, V.~Dagendorf,
  R.~Nolte, U.~Herpers, A.~N. Smirnov, I.~V. Ryzhov, A.~V. Prokofiev,
  P.~Malmborg, D.~Koll\'{a}r, and J.-P. Meulders.
\newblock Excitation functions for the production of radionuclides by
  neutron-induced reactions on {C}, {O}, {Mg}, {Al}, {Si}, {Fe}, {Co}, {Ni},
  {Cu}, {Ag}, {Te}, {Pb}, and {U} up to 180 {MeV}.
\newblock {\em Nuclear Instruments and Methods in Physics Research B},
  343:30--43, 2015.

\bibitem{IAEA-XSECT-ATLAS}
J.~Kopecky, {J.-Ch. Sublet}, J.~A. Simpson, R.~A. Forrest, and D.~Nierop.
\newblock Atlas of neutron capture cross sections.
\newblock Technical report, International Atomic Energy Agency INDC(NDS)-362,
  1997.

\bibitem{PHITS}
Tatsuhiko Sato, Yosuke Iwamoto, Shintaro Hashimoto, Tatsuhiko Ogawa, Takuya
  Furuta, Shin ichiro Abe, Takeshi Kai, Pi-En Tsai, Norihiro Matsuda, Hiroshi
  Iwase, Nobuhiro Shigyo, Lembit Sihver, and Koji Niita.
\newblock Features of particle and heavy ion transport code system ({PHITS})
  version 3.02.
\newblock {\em Journal of Nuclear Science and Technology}, 55(6):684--690,
  2018.

\bibitem{SNS-LESSONS-LEARNED}
Lessons learned from neutron instrument beamline construction.
\newblock Technical report, Oak Ridge National Laboratory --- {SNS} {NFDD}
  {PM-LL-0001-R05}, 2012.

\bibitem{SULLIVAN}
A.~H. Sullivan.
\newblock {\em A guide to radiation and radioactivity levels near high energy
  particle accelerators}.
\newblock Nuclear Technology Publishing ISBN 1 870965 18 3, 1992.

\bibitem{UWE-PRIVATE-COMM}
U.~Filges.
\newblock Private communication, 2021.

\bibitem{PSI-TARGET-NEUTRONICS}
N.~Cherkashyna, D.~DiJulio, T.~Panzner, E.~Rantsiou, U.~Filges, G.~Ehlers, and
  P.~M. Bentley.
\newblock Benchmarking shielding simulations for an accelerator-driven
  spallation neutron source.
\newblock {\em PHYSICAL REVIEW SPECIAL TOPICS---ACCELERATORS AND BEAMS},
  18:083501, 2015.

\bibitem{GEANT4}
S.~Agostinelli et~al.
\newblock Geant4 - a simulation toolkit.
\newblock {\em Nuclear Instruments and Methods A}, 506:250--303, 2003.

\bibitem{SELPH-FAST-NEUTRON-ALBEDO}
W.~E. Selph.
\newblock {\em Weapons Radiation Shielding Handbook}, chapter~4.
\newblock Defense Atomic Support Agency, Washington DC 20301, 1968.

\bibitem{SUPERMIRROR-REFLECTIVITIES}
R.~Kolevatov, C.~Schanzer, and P.~B\"{o}ni.
\newblock Neutron absorption in supermirror coatings: Effects on shielding.
\newblock {\em Nuc. Inst. Meth. Phys. Res. A}, 922:98--107, 2019.

\bibitem{ESS-SOURCE-TERM}
V.~Santoro, D.~DiJulio, P.~Bentley, and L.~Zanini.
\newblock Source term for shielding design of bunker and beamlines at {ESS}.
\newblock Technical report, European Spallation Source {ESS}-0416080, 2017.

\bibitem{JPARC-CURVED-GUIDES}
K.~Niita, K.~Suzuya, K.~Nakajima, R.~Kajimoto, M.~Nakamura, K.~Shibata,
  K.~Soyama, S.~Torii, S.~Harjo, K.~Aizawa, M.~Kawai, T.~Kamiyama, S.~Itoh,
  N.~Torikai, F.~Maekawa, K.~Oikawa, M.~Tamura, M.~Harada, and M.~Arai.
\newblock Neutron beam-line and shield design for neutron scattering
  instruments at {JSNS} {J-PARC} project.
\newblock {\em Proceedings of the Seventeenth Meeting of the International
  Collaboration on Advanced Neutron Sources (ICANS)}, 2: Target
  Stations:640--647, 2005.

\bibitem{ANSELL-PRIVATE-COMM}
S.~Ansell.
\newblock Private communication, 2014.

\bibitem{STUART-ICNS2013}
S.~Ansell.
\newblock Private communication, ICNS Edinburgh, July 2013.

\bibitem{IVERSON-PHOTONUCLEAR}
T.~C. McClanahan, F.~X. Gallmeier, and E.~B. Iverson.
\newblock Photonuclear contributions to {SNS} pulse shapes.
\newblock Technical report, Oak Ridge National Laboratory ---
  {ORNL/TM-2016/758}, 2016.

\bibitem{LILLEY-PHOTO-BG-TALK-PSI}
S.~Lilley, C.~Cazzaniga, and C.~Kane.
\newblock Neutron noise and delayed neutron backgrounds at spallation
  facilities.
\newblock In {\em Efficient Neutron Sources,}, number~14, 2019.

\bibitem{MAHURIN-SNS-DELAYED-NEUTRONS}
R.~Mahurin, G.~Greene, J.~Majewski, H.~Smith, W.~S. Wilburn, D.~Bowman,
  S.~Penttila, L.~Barron, and W.~M. Snow.
\newblock Measurement of delayed neutron production in a tungsten spallation
  neutron target.
\newblock In {\em 2008 APS April Meeting and HEDP/HEDLA Meeting}, volume~53,
  2008.

\bibitem{KIKUCHI-AMATERAS-BACKGROUNDS}
T.~Kikuchi, K.~Nakajima, S.~Ohira-Kawamura, Y.~Inamura, M.~Nakamura, D.~Wakai,
  K.~Aoyama, T.~Iwahashi, and W.~Kambara.
\newblock Background issues encountered by cold-neutron chopper spectrometer
  amateras.
\newblock {\em Physica B: Condensed Matter}, 564:45--53, 2019.

\bibitem{ARGONNE-DELAYED-NEUTRONS}
J.~E. Epperson, J.~M. Carpenter, P.~Thiyagarajan, and B.~Heuser.
\newblock Measurement of the delayed neutron fraction and correction of small
  angle scattering data from a pulsed spallation source.
\newblock {\em Nuclear Instruments and Methods A}, 289:30--34, 1990.

\bibitem{BEWLEY-CADMIUM-PRIVATE}
R.~Bewley.
\newblock Private communication, 2021.

\bibitem{ARAI-TARGET-BACKGROUND}
M.~Arai, L.~Zanini, E.~Bryndt Klinkby, K.~Andersen, R.~Linander, F.~Mezei,
  K.~Niita, M.~Harada, and F.~Maekawa.
\newblock Neutron beam extraction and tailoring useful neutrons to instruments
  at {ESS}.
\newblock {\em Journal of Physics: Conf. Series}, 1021:012035, 2018.

\bibitem{CSPEC-SYSTEM-DESIGN}
P.~Deen, F.~Yamil Moreira, D.~Noferini, W.~Lohstroh, and S.~Longeville.
\newblock {CSPEC} --- system design description.
\newblock Technical report, European Spallation Source ESS-1157621, July 2021.

\bibitem{MEZEI-REP-RATE-MULTIPLICATION}
M.~Russina and F.~Mezei.
\newblock Implementation of repetition rate multiplication in cold, thermal and
  hot neutron spectroscopy.
\newblock {\em Journal of Physics: Conference Series}, 251:012079, 2010.

\bibitem{MAGIC-CONCEPT-DESIGN}
X.~Fabr\`{e}ges.
\newblock Tollgate 2 preliminary system design document for the {MAGiC}
  instrument.
\newblock Technical report, European Spallation Source {ESS}-0131420, 2017.

\bibitem{FILGES-IKON-TALK}
U.~Filges.
\newblock {MAGiC} shielding simulations.
\newblock In {\em IKON Conference, Villigen, Switzerland}, 02 2017.

\bibitem{OVERCOMING-BACKGROUNDS}
N.~Cherkashyna, R.~J. Hall-Wilton, D.~D. DiJulio, A.~Khaplanov, D.~Pfeiffer,
  J.~Scherzinger, C.~P. Cooper-Jensen, K.~G. Fissum, S.~Ansell, E.~B. Iverson,
  G.~Ehlers, F.~X. Gallmeier, T.~Panzner, E.~Rantsiou, K.~Kanaki, U.~Filges,
  T.~Kittelmann, M.~Extegarai, V.~Santoro, O.~Kirstein, and P.~M. Bentley.
\newblock Overcoming high energy backgrounds at pulsed spallation sources.
\newblock {\em Proceedings of the 21st Meeting of the International
  Collaboration on Advanced Neutron Sources ({ICANS-XXI})}, 2014.

\bibitem{COMB-LAYER}
S.~Ansell.
\newblock https://github.com/SAnsell/CombLayer.

\bibitem{STUART-RUBBER-LIFETIME}
N.~Tsapatsaris and S.~Ansell.
\newblock Radiation dose and lifetime calculations for sealing materials used
  at neutron choppers at the {ESS}.
\newblock Technical report, European Spallation Source ESS-0084036, 2018.

\bibitem{SERPENTINE-CONCRETE}
G.~A. {Vasil'ev}, A.~P. Veselkin, Yu.~A. Egorov, V.~A. Kucheryaev, and Yu.~V.
  {Pankrat'ev}.
\newblock Attenuation of reactor radiations by serpentine concrete.
\newblock {\em Atomnaya \'{E}nergiya}, 18:121, 1965.

\bibitem{SERPENTINE-CONCRETE2}
G.~A. Vasil'ev, A.~P. Veselkin, Yu.~A. Egorov, G.~G. Moiseev, and Yu.~V.
  Pankrat'ev.
\newblock Attenuation of pile radiations in serpentinite sand.
\newblock {\em Atomnaya \'{E}nergiya}, 19(4):354--359, 1965.

\bibitem{COLEMENITE-CONCRETE}
K.~Okuno, M.~Kawai, and H.~Yamada.
\newblock Development of novel neutron shielding concrete.
\newblock {\em Nuclear Technology}, 168:545--552, 2009.

\bibitem{COLEMENITE-CONCRETE2}
M.~Kawai, M.~Yonemura, T.~Kamiyama, K.~Okuno, and K.~Niita.
\newblock Development of neutron shielding concrete containing colemanite and
  peridotite.
\newblock {\em JPS Conference Proceedings}, 28:081007, 2019.

\bibitem{CARSTEN-CONCRETE}
D.~D. DiJulio, C.~P. Cooper-Jensen, H.~Perrey, K.~Fissum, E.~Rofors,
  J.~Scherzinger, and P.~M. Bentley.
\newblock A polyethylene-{B}$_4${C} based concrete for enhanced neutron
  shielding at neutron research facilities.
\newblock {\em Nuclear Instruments and Methods in Physics Research A},
  859:41--46, 2017.

\bibitem{BPE-FIRE-TEST}
A.~Bergstrand.
\newblock Fire test according to {EN 13823} ({SBI} method) and {EN ISO}
  11925-2.
\newblock Technical report, SP Technical Research Institute of Sweden (6P09799)
  and {ESS}-0096708, 2017.

\bibitem{BPE-FIRE-TEST2}
A.~Bergstrand.
\newblock Reaction to fire classification report.
\newblock Technical report, SP Technical Research Institute of Sweden
  (6P09799-1) and {ESS}-0096709, 2017.

\bibitem{BPE-PULL-TEST}
J-O. Frederiksen.
\newblock Pull out test of anchors in radiation shielding concrete part 1. test
  of 12 mm hilti anchors.
\newblock Technical report, Danish Technological Institute (703677-1) and
  {ESS}-0099049, 2016.

\bibitem{BPE-PULL-TEST2}
J-O. Frederiksen.
\newblock Pull out test of anchors in radiation shielding concrete part 2. test
  of 20 mm hilti anchors.
\newblock Technical report, Danish Technological Institute (703677-2) and
  {ESS}-0099050, 2016.

\bibitem{DTI-CONCRETE-REPORT}
Thomas~Lennart Svensson and Claus Pade.
\newblock Neutron shielding concrete development of mix design and
  documentation of selected properties.
\newblock Technical report, Danish Technological Institute 654519; European
  Spallation Source {ESS}-0084342, 2015.

\bibitem{MEZEI-BUNKER-DESIGN}
F.~Mezei.
\newblock Revision of bunker: cost efficient shielding design.
\newblock Technical report, European Spallation Source ESS-3528964, 2018.

\bibitem{ESS-0321186}
P.~M. Bentley.
\newblock A study of the 2018 heavy concrete {ESS} bunker.
\newblock Technical report, ESS-0321186, June 2018.

\bibitem{ESTIA-TG3-ESS-0432926}
A.~Glavic, U.~Filges, and A.~Ivanov.
\newblock {ESTIA} --- radation safety analysis.
\newblock Technical report, ESS-432926, 2020.

\bibitem{RAL-WAX-CAN-KNOWLEDGE-TRIP}
D.~Madsen.
\newblock Evaluation of redundant wax tanks at {Rutherford} {Appleton}
  {Laboratory}, {England}.
\newblock Technical report, European Spallation Source {ESS}-0018419, 2014.

\bibitem{PARAFFIN-CAN-FIRE-TEST}
D.~Madsen.
\newblock Evaluation of fire exposed shielding wall.
\newblock Technical report, European Spallation Source {ESS}-0011320, 2014.

\bibitem{STEVENSON-SKYSHINE}
G~R Stevenson and R~H Thomas.
\newblock A simple procedure for the estimation of neutron skyshine from proton
  accelerators.
\newblock {\em Health Physics}, 46:115, 1984.

\bibitem{TARGET-SOURCE-TERM}
G.~Muhrer.
\newblock Handbook assessment of the contribution of the target station to the
  dose at the side boundary through skyshine.
\newblock Technical report, European Spallation Source {ESS}-0065565, 2019.

\bibitem{BUNKER-SOURCE-TERM}
L.~Zanini, D.~DiJulio, V.~Santoro, P.~Bentley, and E.~Klinkby.
\newblock Neutronic design of the bunker wall and roof.
\newblock Technical report, European Spallation Source {ESS}-0416081, 2019.

\bibitem{ACCELERATOR-SOURCE-TERM}
L.~Tchelidze.
\newblock {ESS} accelerator skyshine dose rate maps during normal operations.
\newblock Technical report, European Spallation Source {ESS}-0063681, 2017.

\bibitem{SNS-SURVEY}
Douglas~D. DiJulio, Nataliia Cherkashyna, Julius Scherzinger, Anton Khaplanov,
  Dorothea Pfeiffer, Carsten~P. Cooper-Jensen, Kevin~G. Fissum, Kalliopi
  Kanaki, Oliver Kirstein, Georg Ehlers, Franz~X. Gallmeier, Donald~E.
  Hornbach, Erik~B. Iverson, Robert~J. Newby, Richard~J. Hall-Wilton, and
  Phillip~M. Bentley.
\newblock Characterization of the radiation background at the spallation
  neutron source.
\newblock {\em Journal of Physics: Conference Series}, 746:012033, 2016.

\bibitem{ESBEN-EARTHQUAKE-GAP}
E.~Klinkby.
\newblock Neutronics considerations on dilatation joints.
\newblock Technical report, European Spallation Source {ESS}-0193677, 2017.

\bibitem{BORON-DETECTOR-BACKGROUND}
F.~Piscitelli, G.~Mauri, A.~Laloni, and R.~Hall-Wilton.
\newblock Verification of he-3 proportional counters' fast neutron sensitivity
  through a comparison with he-4 detectors.
\newblock {\em European Physical Journal Plus}, 135:577, 2020.

\end{thebibliography}
\end{document}